\documentclass[letterpaper,aps,prd,twocolumn,tightenlines,preprintnumbers,nofootinbib,showkeys,superscriptaddress]{revtex4-1}
\pdfoutput=1

\usepackage{XCharter}
\usepackage[T1]{fontenc}

\usepackage{fullpage}
\usepackage{amsfonts}
\usepackage{amsmath}
\usepackage{slashed}
\usepackage{amssymb}
\usepackage{graphicx}
\usepackage{epic}
\usepackage{eepic}
\usepackage{epsfig}
\usepackage{latexsym}
\usepackage[dvipsnames]{xcolor}
\usepackage[export]{adjustbox}
\usepackage{float}
\usepackage{multirow}
\usepackage{hyperref}
\usepackage{enumitem}
\usepackage{lipsum}
\hypersetup{colorlinks=true,citecolor=red,linkcolor=NavyBlue,urlcolor=NavyBlue}
\usepackage[caption=false,labelformat=simple]{subfig}
 
\usepackage{natbib}
\usepackage{relsize}
\usepackage[left=1.6cm,right=1.6cm,top=1.8cm,bottom=1.8cm]{geometry}
\usepackage{comment}
\usepackage{shorthand}

\begin{document}
\relscale{1.05}

\title{Discovery prospects of a vectorlike top partner decaying to a singlet boson}

\author{Akanksha Bhardwaj}
\email{akanksha.bhardwaj@glasgow.ac.uk} 
\affiliation{School of Physics \& Astronomy, University of Glasgow, Glasgow G12 8QQ, United Kingdom}

\author{Kartik Bhide}
\email{kartikdbhide17@iisertvm.ac.in}
\affiliation{Indian Institute of Science Education and Research Thiruvananthapuram, Vithura, Kerala, 695 551, India}

\author{Tanumoy Mandal}
\email{tanumoy@iisertvm.ac.in}
\affiliation{Indian Institute of Science Education and Research Thiruvananthapuram, Vithura, Kerala, 695 551, India}

\author{Subhadip Mitra}
\email{subhadip.mitra@iiit.ac.in}
\affiliation{Center for Computational Natural Sciences and Bioinformatics, International Institute of Information Technology, Hyderabad 500 032, India}

\author{Cyrin Neeraj}
\email{cyrin.neeraj@research.iiit.ac.in}
\affiliation{Center for Computational Natural Sciences and Bioinformatics, International Institute of Information Technology, Hyderabad 500 032, India}
\date{\today}


\begin{abstract}
\noindent
The possibility of a vectorlike top partner decaying to a new colourless weak-singlet scalar or pseudoscalar has attracted some attention in the literature. We investigate the production of a weak-singlet charge-$2/3$ $T$ quark that can decay to a spinless boson ($\Phi$) and a top quark at the LHC. Earlier [\href{https://arxiv.org/abs/2203.13753}{2203.13753}], we showed that in a large part of the parameter space, the $T\to t\Phi$ and the loop-induced $\Phi \to gg$ decays become the dominant decay modes for these particles. Here, we investigate the discovery prospects of the $T$ quark in this region through the above decays. In particular, we focus on the $pp\to TT\to (t\Phi)(t\Phi)\to (t(gg))\,(t(gg))$ channel. Separating this signal from the huge Standard Model background is a challenging task, forcing us to employ a multivariate machine-learning technique. We find that the above channel can be a discovery channel of the top partner in the large part of the parameter space where the above decay chain dominates. Our analysis is largely model-independent, and hence our results would be useful in a broad class of new physics models.
\end{abstract}

\maketitle

\section{Introduction}\label{sec:introduction}
\noindent
Various beyond the Standard Model (BSM) scenarios contain heavy companions of the Higgs and the top in the spectrum. For example, we can consider those addressing the hierarchy problem like the partial compositeness models~\cite{Kaplan:1983fs,Kaplan:1991dc,Agashe:2004rs,Ferretti:2013kya,Ferretti:2014qta,Ferretti:2016upr,Banerjee:2022izw}, extra-dimensional models~\cite{Chang:1999nh,Gherghetta:2000qt,Contino:2003ve,Gopalakrishna:2011ef,Gopalakrishna:2013hua,Barcelo:2014kha}, Little-Higgs models~\cite{Arkani-Hamed:2002iiv,Schmaltz:2002wx,Perelstein:2003wd,Martin:2009bg}, etc. Generally, in these models, the top partners ($T$) are vectorlike and decay to a third-generation quark plus a Higgs or a Standard Model (SM) vector boson, $W/Z$. Recently, however, a new exotic decay possibility of $T$ has attracted considerable attention in the literature~\cite{Gopalakrishna:2015wwa,Serra:2015xfa,Anandakrishnan:2015yfa,Banerjee:2016wls,Kraml:2016eti,Dobrescu:2016pda,Aguilar-Saavedra:2017giu,Chala:2017xgc,Moretti:2017qby,Bizot:2018tds,Colucci:2018vxz,Han:2018hcu,Dermisek:2019vkc,Kim:2019oyh,Xie:2019gya,Benbrik:2019zdp,Cacciapaglia:2019zmj,Dermisek:2020gbr,Wang:2020ips,Das:2020ozo,Choudhury:2021nib,Dermisek:2021zjd,Corcella:2021mdl,Dasgupta:2021fzw,Cline:2021iff} where a $T$ quark decays to the top quark along with a scalar ($\phi$) or pseudoscalar ($\eta$) that is SM singlet. This is possible if the singlet boson is lighter than the top partner at least by $m_t$.

In an earlier paper~\cite{Bhardwaj:2022nko}, we explored this possibility. If such a new decay of $T$ exists, the current exclusion limit on $T$ from the direct LHC searches (that assume the $T$ quark decays to SM-only final states) relaxes significantly. In that paper, based on the weak representation of the vectorlike quarks, we obtained some simple generic phenomenological models containing a singlet $\Phi=\{\phi,\eta\}$ and either a weak-singlet vectorlike quark (VLQ) or a $(T\ B)^T$ doublet. For these generic models, we recast the current exclusion limits on $T$. In the singlet $T$ model, $\Phi$ gains coupling with the top quark and the gauge bosons through $T\leftrightarrow t$ mixing after electroweak symmetry breaking (EWSB). This leads to a host of new search possibilities to probe the $T+\Phi$ setup at the LHC.

In that paper, we listed various search channels that can probe different regions of the parameter space.  Prospects of some of these channels at the high luminosity LHC (HL-LHC) have been studied in the literature. For example, Ref.~\cite{Han:2018hcu} examines the prospects of an unusual channel containing six top quarks in the final states. The six top quarks arise from the pair production of $T$ as 
\begin{align}
pp\to T\bar{T}\to (t\Phi)(\bar{t}\Phi)\to (t(t\bar{t}))\,(\bar{t}(t\bar{t})).
\end{align} 
The process is kinematically viable only if $M_\Phi>2m_t$ and $M_T>M_\Phi+m_t$. Reference~\cite{Benbrik:2019zdp} has a detailed analysis of the pair production of the singlet $T$ quark when the top partner decays to a photon pair plus a top quark via $\Phi$: $T\to t\Phi\to t\gm\gm$. This channel is a clean probe of the new decay mode of $T\to t\Phi$ but suffers from low $\Phi\to \gamma\gamma$ branching.

\begin{figure}[!t]
\centering
\includegraphics[width=\columnwidth]{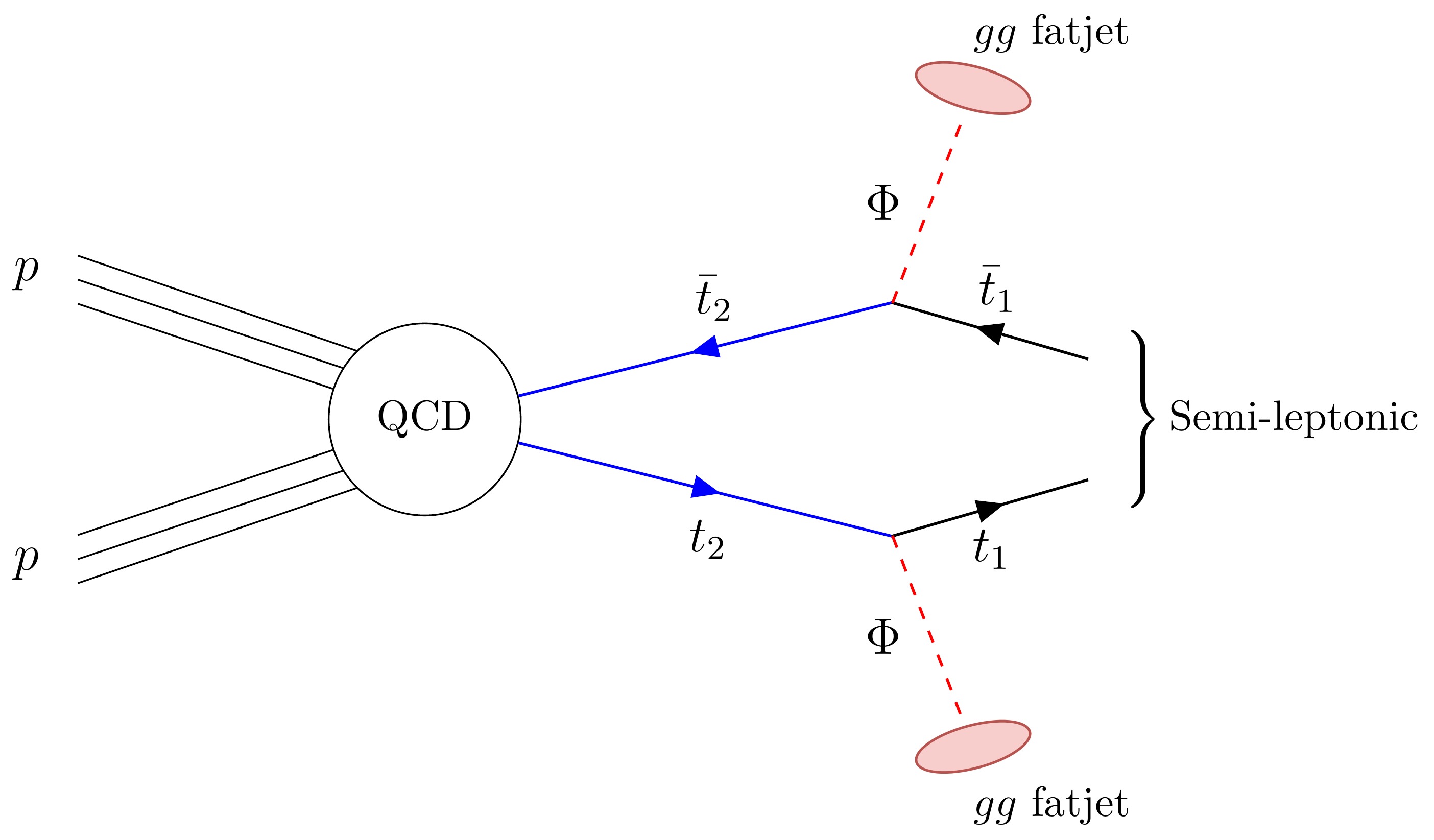}
\caption{The signal topology.}\label{fig:feyn}
\end{figure}

We made two interesting observations in Ref.~\cite{Bhardwaj:2022nko}. First, even though $\Phi$ decays to a $t\bar t$ pair through a tree-level process beyond the mass threshold, it can also decay to a couple of gluons mainly via a $T$ loop. Second, the part of parameter space where the loop-mediated $\Phi$ decay dominates is large, and the corresponding parameters (i.e., couplings) are not fine-tuned. Hence, in a large part of the parameter space, the dominant decay of the $T$ quark is $T\to (t\Phi)\to t(gg)$. In other words, the pair production of $T$ would have the following signature in a large region of the parameter space:
\begin{align}
pp\to T\bar{T}\to (t\Phi)(\bar{t}\Phi)\to (t(gg))\,(\bar{t}(gg)).
\end{align}
(It is not difficult to achieve a large $T\to t\Phi$ branching since the coupling controlling the decay is independent of the small $t\leftrightarrow T$ mixing angle.) Therefore, to probe that parameter region, one has to rely on the above channel (see Fig.~\ref{fig:feyn}). However, since the SM background is huge for this channel, isolating this signal is arduous. In this paper, we take up this task and show that a large parameter region can be discovered through this channel at the HL-LHC with the help of jet substructure and multivariate machine-learning techniques.

This paper is organised as follows: In Section~\ref{sec:model_description} we describe the singlet $T$ model from Ref.~\cite{Bhardwaj:2022nko} and the constraints on its parameters, in Section~\ref{sec:analysis} we describe our analysis from event generation to multivariate analysis and present the results, and conclude in Section~\ref{sec:conclusion}.

\section{The singlet \texorpdfstring{$T$}{T} model}\label{sec:model_description}
\noindent
From Ref.~\cite{Bhardwaj:2022nko}, we select the simple extension of the SM containing a TeV-range weak-singlet VLQ, $T\equiv(\textbf{3},\textbf{1},2/3)$ and a lighter SM-singlet scalar or pseudoscalar. We can write the top-sector mass terms in the interaction basis as (more details are given in the Appendix):
\begin{equation}  \mathcal{L}\supset
    \left(\begin{array}{cc}
    \bar{t}_{L} & \bar{T}_L\end{array}\right)\left(\begin{array}{cc}
    m_{t} & \mu_{1}\\ 0 & M_{T}
    \end{array}\right)\left(\begin{array}{c}
    t_{R}\\ T_{R}
    \end{array}\right)+\text{h.c.}
\end{equation}
Throughout the paper, we use similar notations as Ref.~\cite{Bhardwaj:2022nko}. We diagonalise the mass matrix by biorthogonal rotations through two mixing angles $\theta_{L}$ and $\theta_R$,
\begin{equation} \left(\begin{array}{c}
    t_{L/R}\\ T_{L/R}
    \end{array}\right)=\left(\begin{array}{cc}
    c_{L/R} & s_{L/R} \\ -s_{L/R} & c_{L/R}
    \end{array}\right)\left(\begin{array}{c}
    t_{1_{L/R}}\\ t_{2_{L/R}}
    \end{array}\right)
\end{equation}
where $s_P=\sin\theta_P$ and $c_P=\cos\theta_P$ for the two chirality projections, and $t_1$ and $t_2$ are the mass eigenstates. We identify the $t_1$ quark with the physical top quark (we refer to it just as $t$ for simplicity). For a small $t\leftrightarrow T$ overlap, $t_2$ is mostly the $T$ quark. The top partner can decay to the SM quarks along with a $W$, $Z$, or $h$ boson. With the introduction of $\Phi$, a new decay mode opens up: $t_2\to t\Phi$. For this new decay, the relevant terms in the Lagrangian are:
\begin{equation}
\mc{L}_{int}^{T} = - \lm_{\Phi}^a\Phi\,\bar{T}_L\Gm T_R - \lm_{\Phi}^b\Phi\,\bar{T}_L\Gm t_R   + {\tr h.c.}
\end{equation}
where $\Gm=\lt\{1,i\gm_5\rt\}$ for $\Ph =\lt\{\ph,\et\rt\}$. Expanding $t$ and $T$ in terms of $t_{1}$ and $t_{2}$ gives:
\begin{align}
\mc{L} \supset & - \lm_{\Ph}^a\Ph \lt( c_L\bar{t}_{2L}-s_L\bar{t}_{L}\rt)\Gamma\lt(c_R {t}_{2R}-s_R {t}_{R}\rt) \nn\\
                      & -\lm_{\Ph }^b \Ph \lt(c_L\bar{t}_{2L} - s_L\bar{t}_{L}\rt)\Gamma\lt(c_R {t}_{R} + s_R {t}_{2R}\rt) + {\tr h.c.}\label{eq:LagintS}
\end{align}

Before EWSB, $\Phi$ essentially couples only with the top partner. When the symmetry breaks, $T$ mixes with the SM top quark, and through it, $\Phi$ couples to the top quark. It also has loop-mediated effective couplings to the SM gauge bosons. Thus it can decay to $t\bar t$, $gg$, $\gm\gm$, $Z\gm$, $ZZ$ final states (when kinematically allowed). The expressions for the decay widths in the dominant decay modes of $\Phi$ are available in Ref.~\cite{Bhardwaj:2022nko}. As explained in the Introduction, in this paper, we consider the $\Phi$ boson dominantly decaying to a pair of gluons.
\medskip

\begin{figure}[!t]
\centering
\includegraphics[scale=0.75]{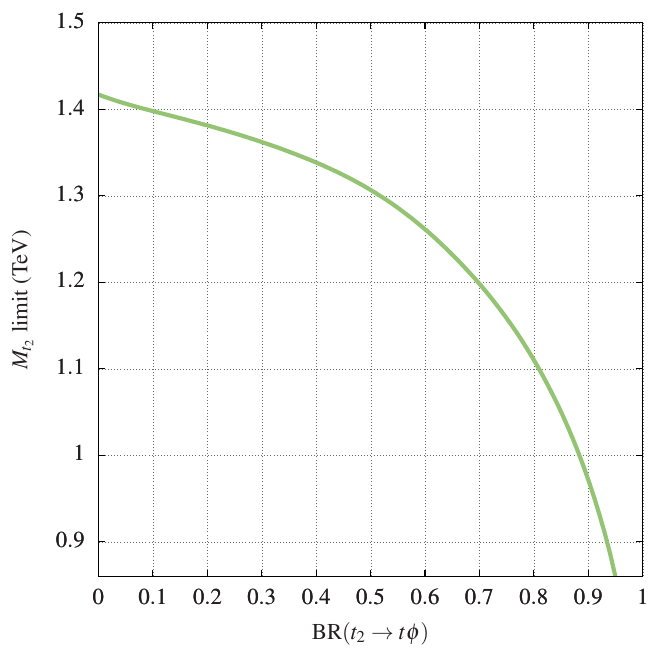}
\caption{LHC exclusion limits on $t_2$ in the singlet $T$ model as a function of the branching ratio in the new decay mode.}
\label{fig:t2_LHClimits}
\end{figure}

\noindent 
{\bf Exclusion bounds on ${\mathrm T}$:} The VLQ searches at the LHC assume that a top partner can only decay to a third-generation quark and a SM boson. Depending on the weak representations of $T$ and branching ratios (BRs) in the conventional decay modes, the limits from the LHC searches vary from $1.29$--$1.57$ TeV~\cite{ CMS:2019eqb,ATLAS:2021ibc,ATLAS:2022ozf,ATLAS:2021ddx}. Currently, the mass limit on a weak-singlet $T$ stands at $1.29$ TeV~\cite{ATLAS:2021ibc}. With the introduction of the singlet $\Phi$, the assumption on the $\Phi$ decays changes to
\begin{equation}
     \beta_{b W} + \beta_{tZ} + \beta_{th} = 1 - \beta_{t\Phi},
     \label{br_constraint}
\end{equation}
where $\beta_{X}$ is the BR for the $t_2 \to X$ decay. For a heavy $t_2$, $\beta_{bW} \approx 2\beta_{tZ} \approx 2\beta_{th}$ in the singlet $T$ model. With this, one can recast the  bounds from the latest searches~\cite{ATLAS:2021ibc, CMS:2020ttz, CMS:2019eqb} in each final state and pick the strongest limit. In Fig.~\ref{fig:t2_LHClimits}, we show the recast mass limits on $T$ in the presence of the new decay mode from Ref.~\cite{Bhardwaj:2022nko}. 
\medskip

\begin{figure*}
\captionsetup[subfigure]{labelformat=empty}
\subfloat[\quad\quad(a)]{\includegraphics[width=0.95\columnwidth]{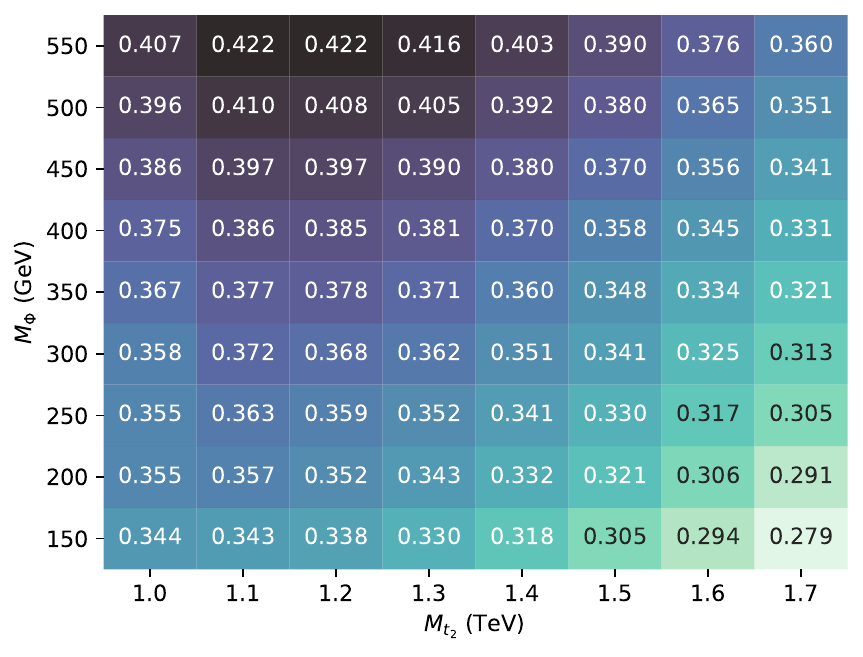}\label{fig:signal_cuteff}}\hspace{0.5cm}
\subfloat[\quad\quad(b)]{\includegraphics[width=0.95\columnwidth]{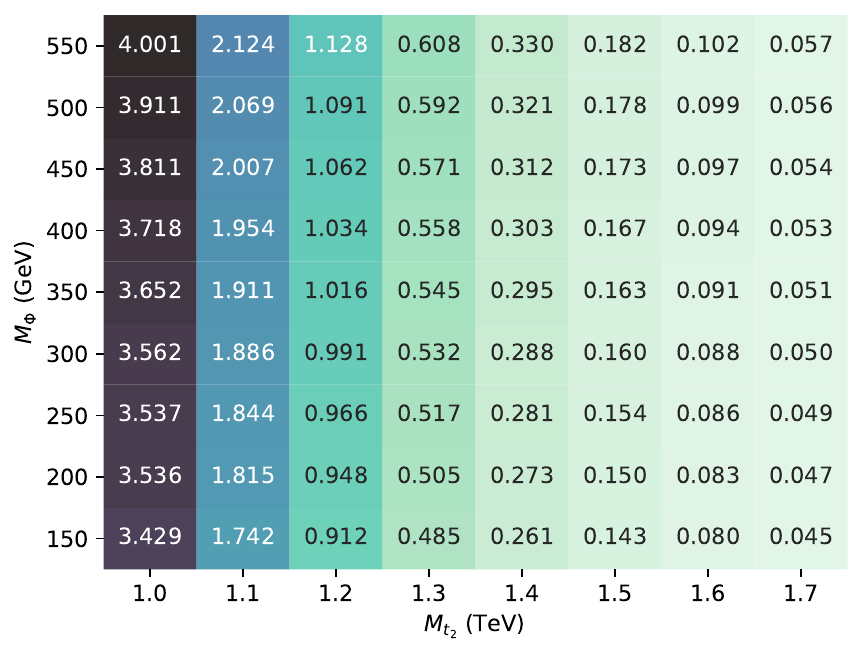}\label{fig:signal_finalxsec}}
\caption{(a) Selection efficiency for the signal process and (b) effective cross section in $fb$ for the signal process, i.e., $\sigma_{\text{gen}} \times \epsilon \times K $ where $\sigma_{\text{gen}}$ refers to the leading order (generation-level) cross section, $\epsilon$ is the selection efficiency and $K$ is the $K$ factor.}
\label{fig:sig_eff}
\end{figure*}

\noindent 
{\bf Bounds on ${\mathrm \Phi}$:} In this paper, we are primarily concerned about the BR of the $\Phi\to gg$ decay. This decay is mediated by $t$ and $t_2$ loops in the physical basis. As mentioned earlier, this decay is dominant over a large region of phase space allowed by the LHC data.\footnote{In Ref.~\cite{Bhardwaj:2022nko}, we have put bounds on the $\Phi gg$ coupling from the ATLAS resonance search data in the diphoton mode~\cite{ATLAS:2021uiz}.} In other words, we can choose a value of BR($\Phi\to gg$) and find suitable parameter combination(s) from the allowed region without any fine-tuning (for both $\phi$ and $\eta$). 

\section{Collider Analysis}\label{sec:analysis}
\noindent
We use {\sc MadGraph5\_aMC@NLOv3.2.0}~\cite{Alwall:2014hca} to generate the signal and background events at the $14$ TeV LHC with {\sc CTEQ6L1} parton distribution functions~\cite{Pumplin:2002vw},  {\sc Pythia8v8.2}~\cite{Sjostrand:2014zea} for parton showering and hadronisation, and {\sc Delphesv3.5.0}~\cite{deFavereau:2013fsa} to simulate the detector environment. We use the default CMS card with  jets of  radius $R=0.4$ clustered with the anti-$k_{t}$ algorithm~\cite{Cacciari:2008gp} in {\sc FastJet v3.3.4}~\cite{Cacciari:2011ma}. For the higher-order effects, we multiply the leading order cross sections with the highest order $K$-factors available in the literature. For the signal, we estimate the NNLO $K$-factor as $1.43$ in the mass range of interest using the {\sc Hathor} package~\cite{Aliev:2010zk}. 

\begin{table*}
\begin{centering}
\begin{tabular*}{\textwidth}{l @{\extracolsep{\fill}} rlrr}
\hline 
\multirow{2}{*}{Process} & Selection cut & \multirow{2}{*}{Estimated $K$-factor} & Eff. cross-sec. $\sg_{\rm eff}=$ & Events at 3 ab$^{-1}$\tabularnewline
 & efficiency $\left(\epsilon\right)$ &  & $\sigma_{\text{gen}}\times\epsilon\times K$ (fb) & $\sigma_{\text{eff}}\times \mc L$\tabularnewline
\hline 
\hline 
$t_{l}\bar{t}_{h}\ (+2j)$ & $0.028759$ & $1.783$ (N$^{3}$LO) \cite{Muselli:2015kba} & $838.31$ & $2,514,940$\tabularnewline
$W_{l}\ (+2j)$ & $0.003657$ & $1.189$ (NLO) \cite{Balossini:2009sa} & $57.41$ & $172,222$\tabularnewline
$t_{l}\bar{t}_{l}\ (+2j)$ & $0.014806$ & $1.783$ (N$^{3}$LO) \cite{Muselli:2015kba} & $29.31$ & $87,918$\tabularnewline
$t_{l}+j\ (+2j)$ & $0.018136$ & $1.175$ (N$^{2}$LO) \cite{Kidonakis:2015nna} & $18.01$ & $54,027$\tabularnewline
$1$ lept. $tW\ (+2j)$ & $0.018806$ & $1.333$ (aN$^{2}$LO) \cite{Kidonakis:2015nna} & $12.58$ & $37,747$\tabularnewline
$1$ lept. $t\bar{t}W\ (+j)$ & $0.027047$ & $1.572$ (NLO$+$N$^{2}$LL) \cite{Broggio:2019ewu} & $6.14$ & $18,411$\tabularnewline
$t_{l}\bar{t}_{h}H_{b}\ (+j)$ & $0.030843$ & $1.410$ (NLO$+$N$^{2}$LL) \cite{Broggio:2019ewu} & $4.37$ & $13,125$\tabularnewline
$t_{l}\bar{t}_{h}Z_{h}\ (+j)$ & $0.023209$ & $1.662$ (NLO) \cite{LHCHiggsCrossSectionWorkingGroup:2011wcg} & $3.53$ & $10,598$\tabularnewline
\hline
\multicolumn{1}{c}{} & \multicolumn{1}{c}{} & \multicolumn{1}{c}{} &  & \textbf{$2,908,998$}\tabularnewline
\cline{5-5} 
\end{tabular*}
\par\end{centering}
\caption{Selection efficiency and the expected number of events at $3$ ab$^{-1}$ for the background processes. Here $\sigma_{\text{gen}}$ refers to the leading order (generation-level) cross section after the $H_{T}>500$ GeV cut. The subscripts $l,h$ refer to the leptonic and hadronic decay modes. The additional jets (in parentheses) indicate the number of matched hard jets, where the jet-parton shower matching is done using the MLM prescription~\cite{Mangano:2006rw}.}\label{table:bkg_eff}
\end{table*}
\begin{figure*}
\captionsetup[subfigure]{labelformat=empty}
\subfloat[\quad\quad(a)]{\includegraphics[width=0.95\columnwidth]{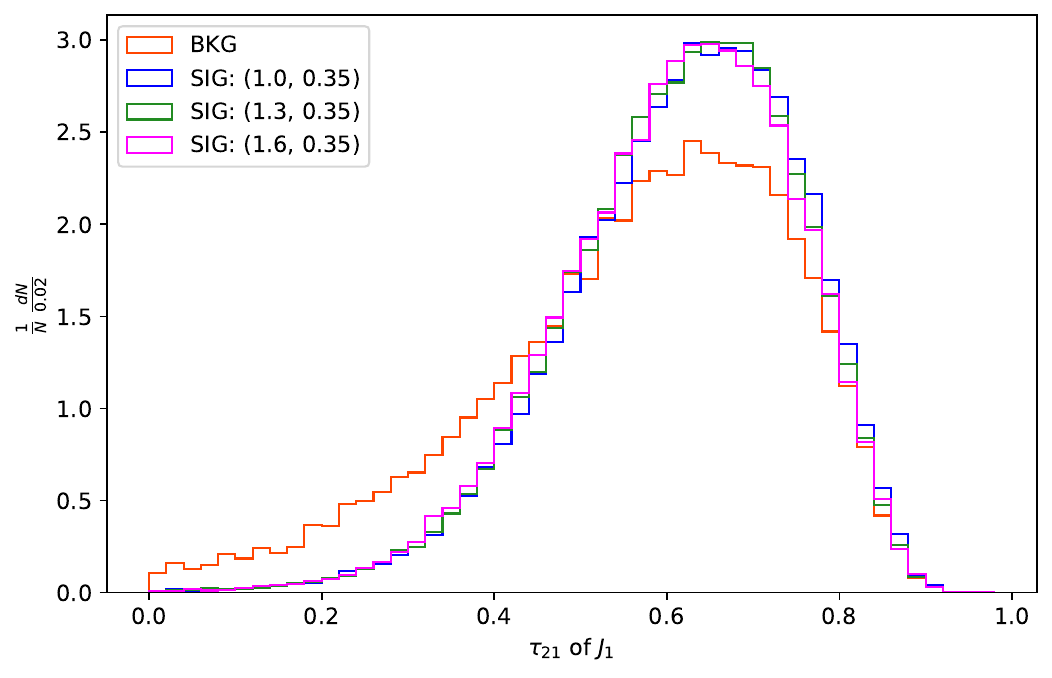}\label{fig:tau21_distrib}}\hspace{0.5cm}
\subfloat[\quad\quad(b)]{\includegraphics[width=0.95\columnwidth]{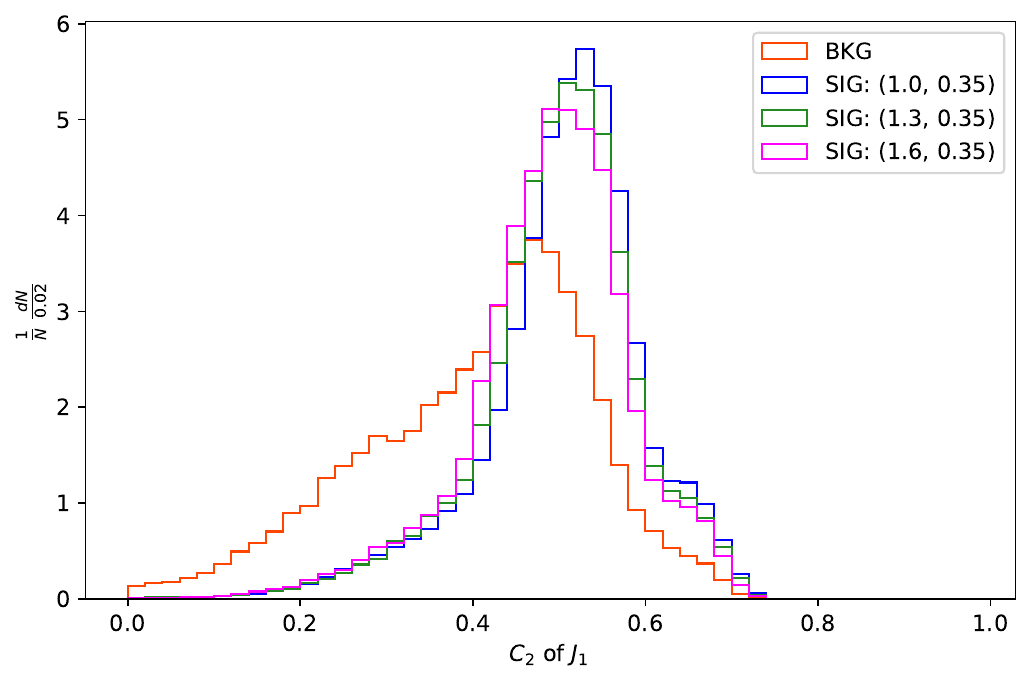}\label{fig:c2_distrib}}
\caption{The features (a) $\tau_{21}$ and (b) $C_{2}$ calculated for the leading $p_T$ fatjet for different $\left(m_{t_2},m_{\phi}\right)$ mass points}
\label{fig:sig_bg1}
\end{figure*}

\subsection{Pair production: Signal topology and selection criteria}\label{sec:sigtop}
\noindent 
The signal process is $pp\rightarrow t_2\bar{t}_2 \rightarrow \Phi t\, \Phi \bar{t} $, with each of the $\Phi$'s decaying to couple of gluons and one of the top quarks decaying leptonically and the other, hadronically. For simulating the signal events, we set  $BR\left(T\rightarrow t\Phi \right)=BR\left(\Phi\rightarrow gg\right)=1$. For our analysis, we scan over the mass ranges $M_{t_2}\in \left[1.0, 1.7 \right]$ TeV in steps of $100$ GeV and $M_{\Phi} \in \left[150, 550\right]$ GeV in steps of $50$ GeV. For each mass point in the scan, we choose the parameters such that the narrow-width approximation is valid for both $t_2$ and $\Phi$. So, from here on, we  treat the BRs as free parameters (rather than the couplings) as we are interested in the decays. This also makes our results interpretable in terms of both $\phi$ and $\eta$, even though we only use the scalar $\phi$ to generate the signals for our analysis.
When $M_{t_2}\gg M_\Phi$, the singlet produced from a $t_2$ decay would be boosted and produce a fatjet. Based on the signal topology ($1$ lepton, $2$ $b$-jets, $2$ $\Phi$-jets) we design the following selection criteria:
\begin{enumerate}
    \item \emph{Exactly $1$ lepton in the event (either $e$ or $\mu$) with $p_{T}>10$ GeV and $\left|\eta \right|<2.5$}. To be accepted as a lepton, the separation between a lepton candidate (identified from the tracks) and its nearest AK4 jet\footnote{AK4 jets are clustered with the anti-$k_{T}$ algorithm~\cite{Cacciari:2008gp} with radius $R=0.4$.} should be greater than $\Delta R=0.4$. 
    
    \item \emph{The scalar sum of the $p_{T}$'s of the AK4 jets, $H_{T}>900$ GeV}.
    
    \item \emph{At least $2$ $b$-tagged jets in the event,} where $b$-tagging is done with the default Delphes module~\cite{CMS:2012feb}. We compare the efficiency of the default Delphes module with the DeepCSV algorithm~\cite{CMS:2017wtu} at the medium working point and find that the tagging efficiencies of the two algorithms in $p_{T}$ range of our interest are comparable. We call the leading and sub-leading $p_{T}$ $b$-jets as $b_{1}$ and $b_{2}$, respectively.
    
    \item \emph{At least $2$ fatjets in the event.} The candidate fatjets are clustered from the calorimeter towers with the Cambridge-Aachen clustering algorithm~\cite{Dokshitzer:1997in} and have $R=1.2$ and $p_{T}>300$ GeV. The candidates are then groomed with the SoftDrop~\cite{Larkoski:2014wba} algorithm for $z_{\rm cut}=0.1$ and $\beta=0.2$. If a candidate fatjet has at least $3$ constituent hadrons needed to compute the $3$-point Energy Correlation Functions, it is accepted as a fatjet. We denote the leading and sub-leading $p_T$ fatjets as $J_{1}$ and $J_{2}$, respectively.
\end{enumerate}
A summary of the cross sections, selection efficiencies and expected number of events at $3$ ab$^{-1}$ for the signal is shown in Fig~\ref{fig:sig_eff}.

The efficiency trends seen in Fig \ref{fig:signal_cuteff} can be explained by the boost of the final state. We see an increase followed by a decrease in selection efficiency for a fixed $M_{\Phi}$. This is the result of an interplay between selection criteria: as $M_{t_2}$ increases, the $H_{T}$ cut becomes more efficient, but the lepton-number and $b$-tagging cuts  worsen due to poor isolation and reduction in efficiency in the boosted regime. In contrast, for a fixed $M_{t_2}$, the selection  efficiency monotonically increases. As $M_{\Phi}$ increases, the boost available to the top quark reduces, so the $b$-jet is easily tagged and the lepton is more isolated. While a slight reduction in the efficiency of the fatjet cut is seen with increasing $M_{\Phi}$ (due to subjets not being fully resolved in a large-$R$ jet as a consequence of the high mass scalar having lower boost), the overall increase in selection efficiency is governed by the improvement in $b$-tagging.

\subsection{Background processes}
\noindent
Based on the selection criteria, the significant background processes we consider are the semileptonic and leptonic $t\bar{t}$ production, $W+$jets, single top production via $tj$ and $tW$ processes, and $t\bar{t}X$ production where $X=\{W,Z,H\}$. Of these, the semileptonic $t\bar{t}$ process is the most dominant, followed by the sizeable contributions from the $W+$jets, leptonic $t\bar{t}$, and single top processes. In order to reduce computation time during the background event generation, we employ a strong generation-level cut: $H_{T}>500$ GeV. After passing through the selection cuts, the contribution of the processes like $Z+$jets, di-boson and tri-boson productions become negligible. The background cross sections, selection efficiencies and expected number of events at $3$ ab$^{-1}$ are shown in Table~\ref{table:bkg_eff}.

\begin{figure*}
\centering
\subfloat[]{
\includegraphics[width=0.95\columnwidth]{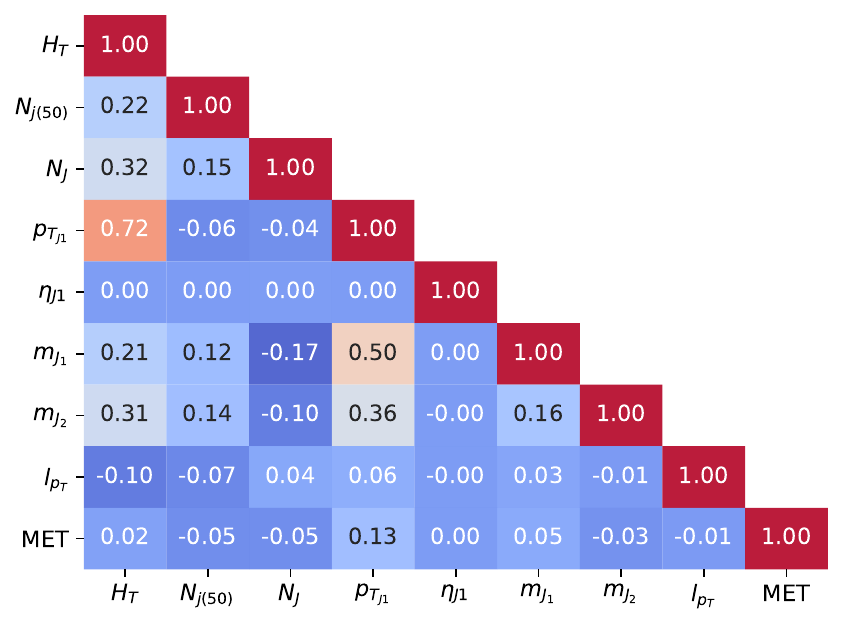}
}\hspace{0.5cm}\subfloat[]{
\includegraphics[width=0.95\columnwidth]{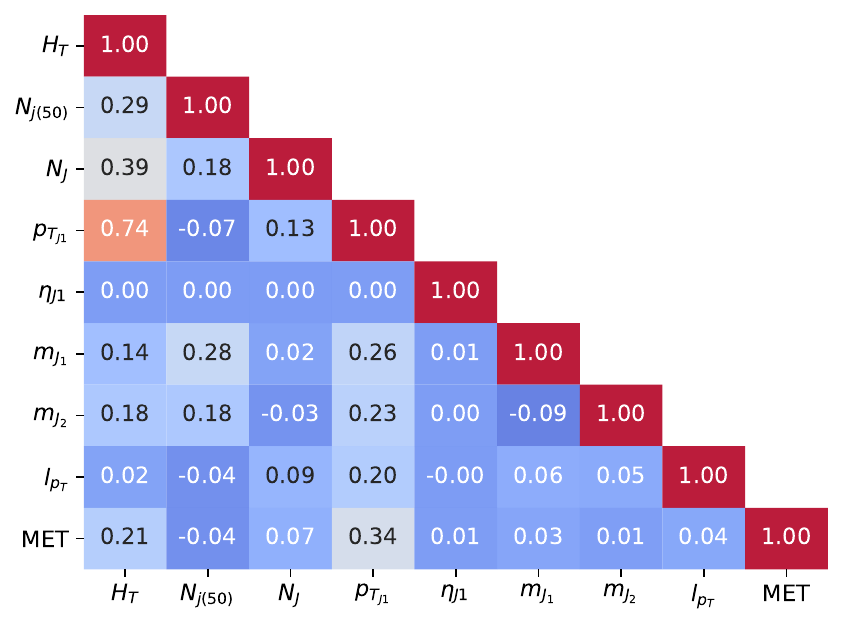}
}
\caption{Linear correlation coefficients for (a) signal and (b) background between the inputs chosen for multivariate analysis at the benchmark mass point. Note that the scale goes from $+1$ to $-0.2$. }\label{fig:correlation_matrices}
\end{figure*}

\subsection{Multivariate Analysis}\label{subsec:mva}
\noindent
We perform a multivariate signal-vs-background discrimination analysis with the Boosted Decision Tree (BDT) algorithm.
\medskip

\noindent
{\bf Choice of the input features:} There are $6$ physical objects from which we can extract information, namely the lepton, the missing energy, the $2$ leading $b$-tagged jets and the $2$ leading fatjets. To determine the best features as inputs to the BDT, we start with a large set split into some groups:
\begin{enumerate}
    \item \emph{Measure of the physical objects:} the number of $b$-tagged jets, fatjets, and the number of jets with $p_{T}$ over $20$, $40$, $50$ and $60$ GeV. We consider these jet-number features so that the high $p_{T}$ jets from the $\Phi$ decay are captured. This group also includes the $H_{T}$ feature.
    
    \item \emph{Basic kinematic variables:} This group includes  $p_{T}$, $\eta$, and $\Phi$ of the $6$ objects considered. We also include the energy $E$ of the visible $5$ objects.
    
    \item \emph{Separation variables:} We consider $\Delta\eta$, $\Delta\phi$ and $\Delta R=\left(\Delta\eta^{2}+\Delta\phi^{2}\right)^{1/2}$ among the $6$ objects. In total, there are $3\times~ C^{\hspace{-2.5ex}6}_{2}=45$ features in this group.
    
    \item \emph{Fatjet features:} For the first two leading $p_T$ fatjets, we consider their masses and the number of constituents. Additionally, we include the well-known jet substructure features of the two fatjets, namely NSubjettiness~\cite{nsubjettines} $\tau_{1}$, $\tau_{2}$, $\tau_{3}$, $\tau_{21}$ and $\tau_{32}$ with $\beta=0.5$ and the Energy Correlation Function \cite{ecf_features} features ECF$_{1}$, ECF$_{2}$, ECF$_{3}$ and $C_{2}$ with $\beta=0.2$.
    
    \item \emph{Reconstructed $T$ mass:} One can reconstruct the transverse mass of $T$ through the semileptonic decay products as $M_{t_2}^{T}=m\left(l\nu b J\right)$ if one assigns the missing energy to the neutrino. We calculate $4$ such reconstructed masses for the $2$ $b$-tagged jets and $2$ fatjets and choose the ones with the minimum and maximum masses.
\end{enumerate}

By looking at the above features for the generated signal and background events, we narrow down the list by discarding objects with less than $1$\% method unspecific separation, defined in Ref.~\cite{Hocker:2007ht} as:
\begin{equation}
    \left<S^{2}\right>=\frac{1}{2}\int dy \frac{\left(\hat{y}_{S}\left(y\right)-\hat{y}_{B}\left(y\right)\right)^{2}}{\hat{y}_{S}\left(y\right)+\hat{y}_{B}\left(y\right)}\label{eqn:tmva_separation}
\end{equation}
where $\hat{y}_{S}$ and $\hat{y}_{B}$ are the probability density functions of the signal and background respectively for a particular feature $y$. This quantity is equal to zero for identical signal and background shapes and $1$ for shapes with no overlap. Naturally, features with higher method unspecific separation are better at discriminating between signal and background events. We also reject features like (most of) the separation variables that are highly correlated with others or do not perform well for different BDT parameter settings.  A large number of basic kinematic features and the reconstructed masses are also discarded for the same reasons. Among the measures of physical objects, we retain $H_{T}$, the number of jets with $p_{T}>50$ GeV (since this one has the best separation compared to the other jet count features) and the number of fatjets.

In the signal, we have two boosted $\Phi$-jets that are fatjets basically made up of two gluons. Simple variables such as $N_\text{consti}$ i.e. the number of (charged) hadrons in the fatjet are known to be good discriminators between $q$- and $g$-initiated jets~\cite{CMS-PAS-JME-13-002}. However, since it is highly correlated with the fatjet mass, we don't consider it. The traditional observables like $\tau_{21}$ and $C_{2}$ hint at the $2$-pronged nature of the signal fatjets but similar behaviour is also seen in the background fatjets since the dominant semileptonic $t\bar{t}$ background has both boosted $W$-jet and a $3$-pronged $t$-jet. As a result, the distributions of these observables show little separation between the signal and background (see Fig.~\ref{fig:sig_bg1}). In addition, we find that the simpler observables  $\tau_{1,2,3}$ and $\text{ECF}_{1,2,3}$ show a slight separation but are correlated with each other and with the fatjet masses and $p_{T}$'s. Hence, we do not include any jet substructure observables in our analysis.
\medskip

\noindent
{\bf The final list:}
The final list of features chosen as inputs to the boosted decision tree algorithm are:\footnote{During preliminary analysis, we found the Lund Jet Plane to be a promising tool in tagging the $gg$ fatjets, similar to the $H\rightarrow gg$ tagging done in Ref.~\cite{Khosa:2021cyk}. However, it requires complex neural network machinery beyond the scope of this paper.}
\begin{itemize}
    \item $H_{T}$, the scalar sum of $p_{T}$'s of all final state hadrons.
    \item $N_{j\left(50\right)}$ and $N_{J}$, the number of AK4 jets with $p_{T}>50$ GeV and the number of fatjets respectively.
    \item $m_{J_1}$ and $m_{J_2}$, the masses of the leading and sub-leading fatjets respectively.
    \item $p_{T_\ell}$ and MET, the $p_{T}$ of the lepton and the transverse missing energy respectively.
    \item $p_{T_{J1}}$ and $\eta_{J1}$, the $p_{T}$ and $\eta$ coordinate of the leading fatjet respectively.
\end{itemize}
In Fig.~\ref{fig:correlation_matrices}, we show the linear correlation coefficients between the input features, defined as:
\begin{equation*}
    \rho\left(X,Y\right)=\frac{\left<XY\right>-\left<X\right>\left<Y\right>}{\sigma_{X}\sigma_{Y}}
\end{equation*}
where $\langle A\rangle$ and $\sigma_{A}$ denote the expectation value and standard deviation respectively for a one-dimensional dataset $A$.

\begin{figure*}
\centering
\subfloat[]{\includegraphics[width=0.32\textwidth]{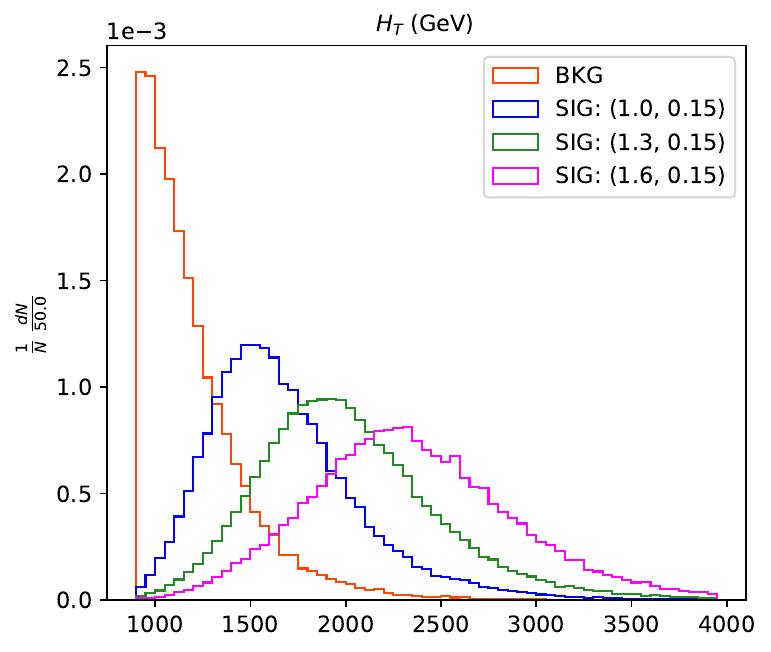}}\quad
\subfloat[]{\includegraphics[width=0.32\textwidth]{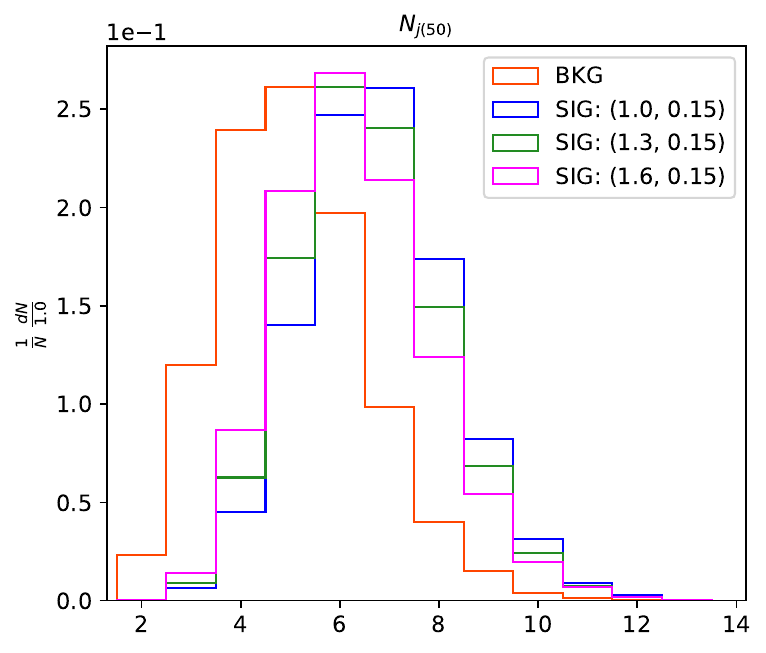}}\quad
\subfloat[]{\includegraphics[width=0.32\textwidth]{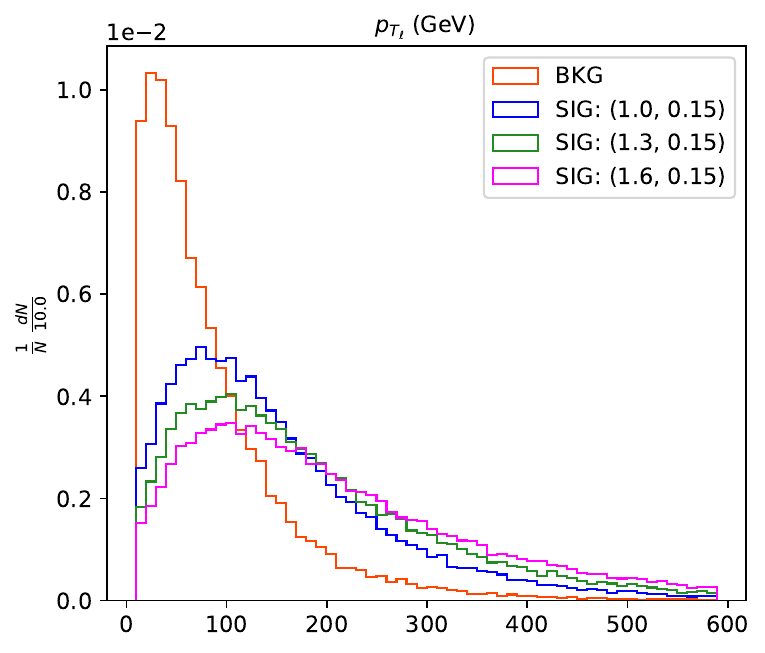}}\\
\subfloat[]{\includegraphics[width=0.32\textwidth]{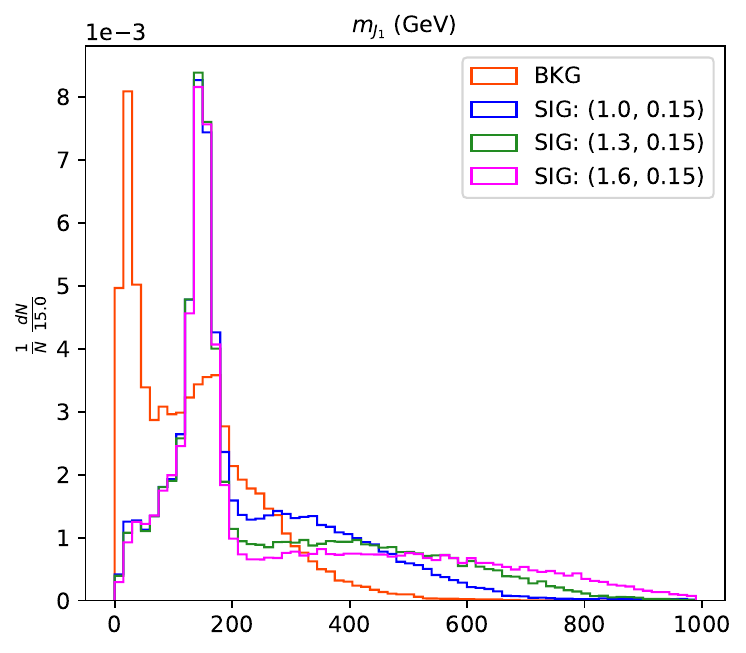}}\quad
\subfloat[]{\includegraphics[width=0.32\textwidth]{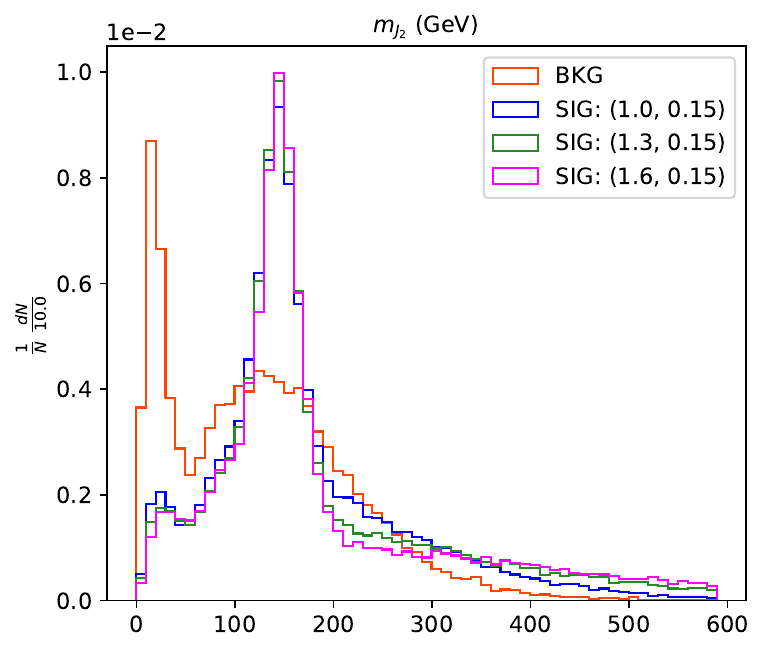}}\quad
\subfloat[]{\includegraphics[width=0.32\textwidth]{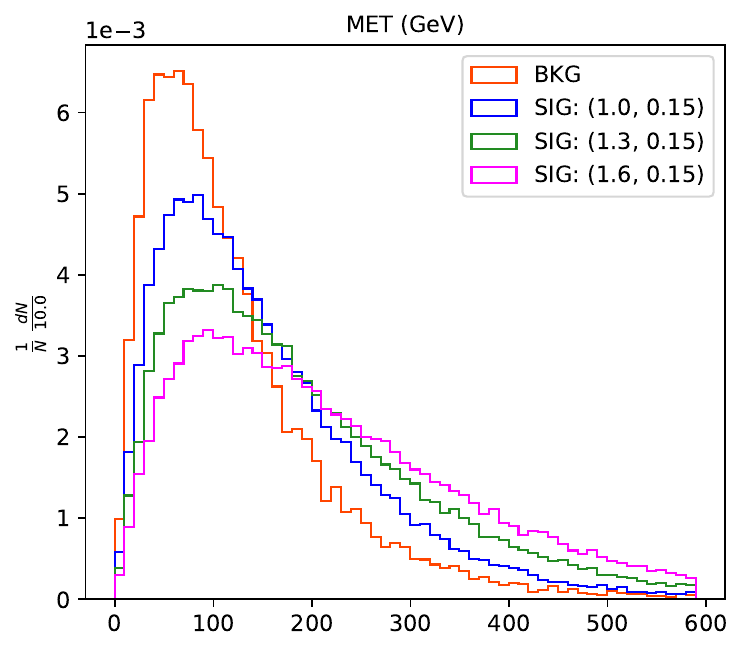}}\\
\subfloat[]{\includegraphics[width=0.32\textwidth]{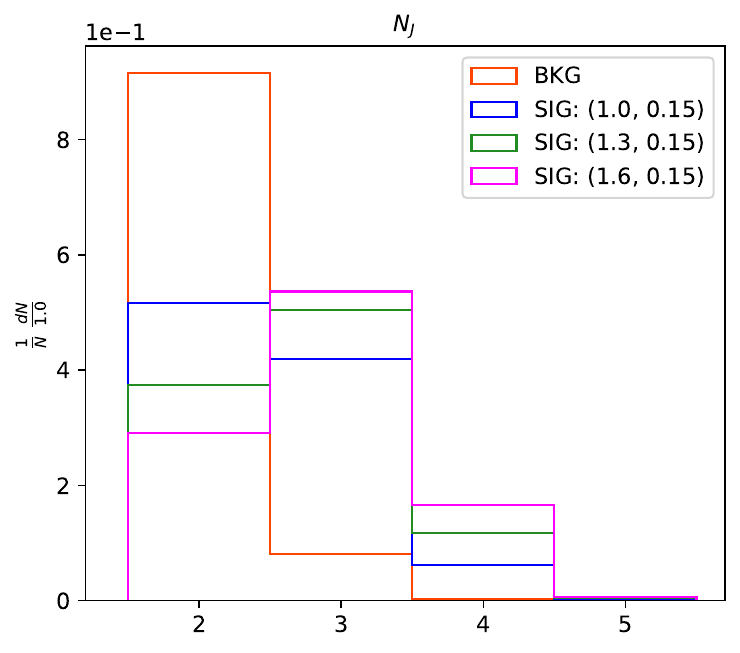}}\quad
\subfloat[]{\includegraphics[width=0.32\textwidth]{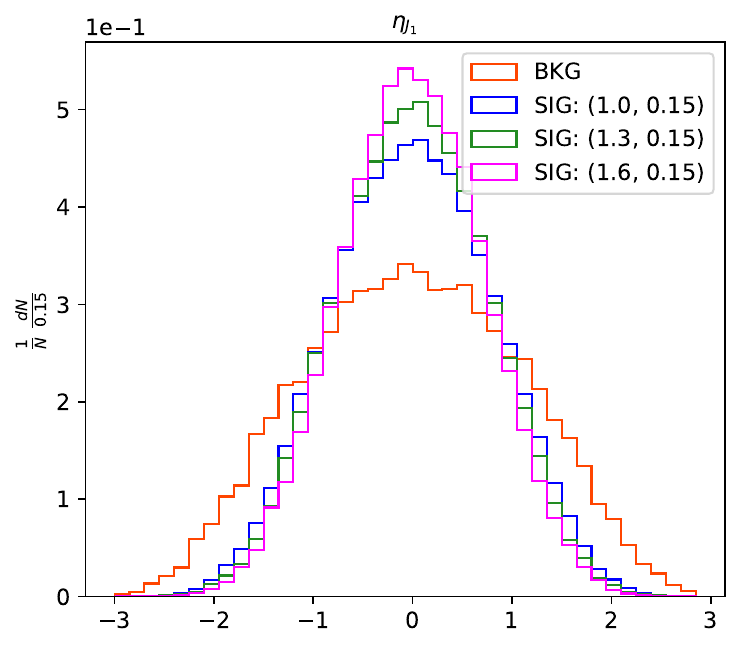}}\quad
\subfloat[]{\includegraphics[width=0.32\textwidth]{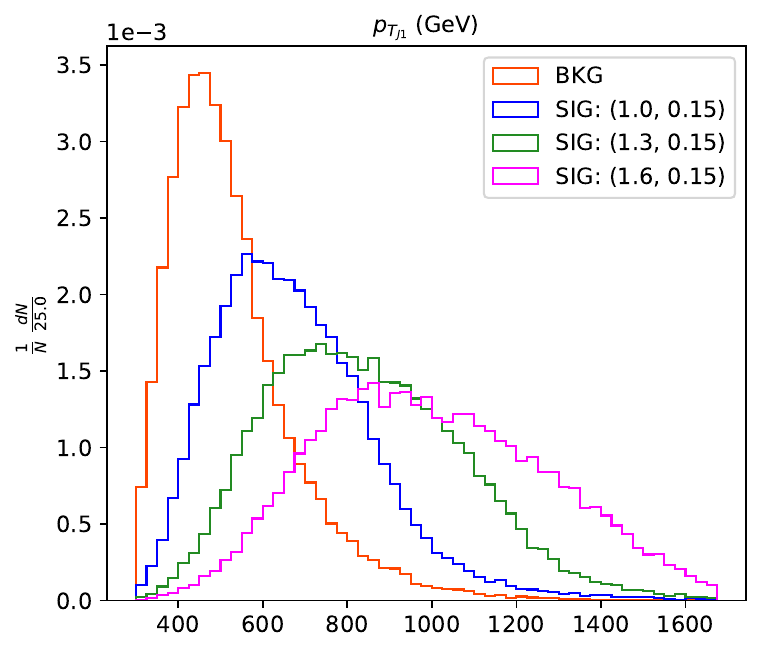}}
\caption{Normalised input feature distributions for signal and background at a fixed $M_{\Phi}=150$ GeV and for three different $M_{t_2}$ points: $M_{t_2}=1.0$, $1.3$ and $1.6$ TeV. The background distribution includes all the backgrounds processes mentioned in the text.}\label{fig:inputvars_mphi150}
\end{figure*}
\begin{figure*}
\centering
\subfloat[]{\includegraphics[width=0.32\textwidth]{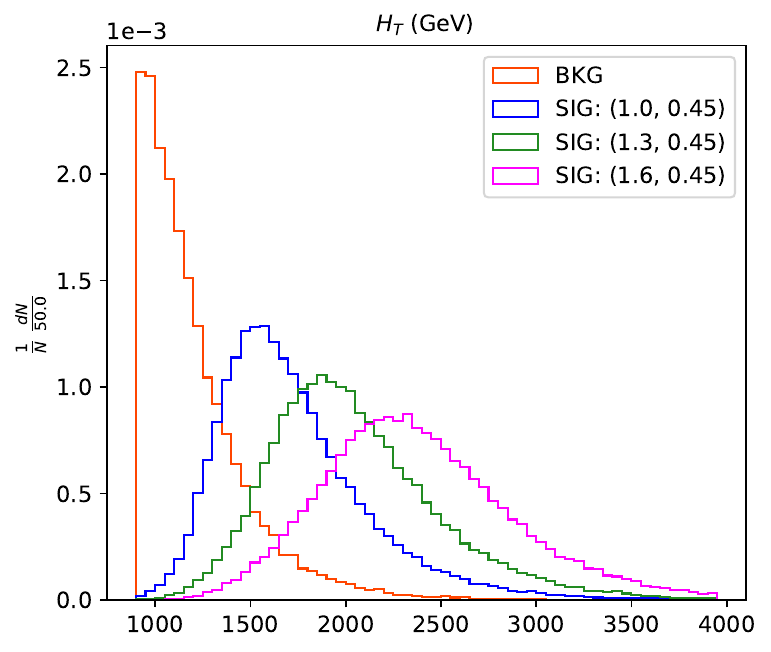}}\quad
\subfloat[]{\includegraphics[width=0.32\textwidth]{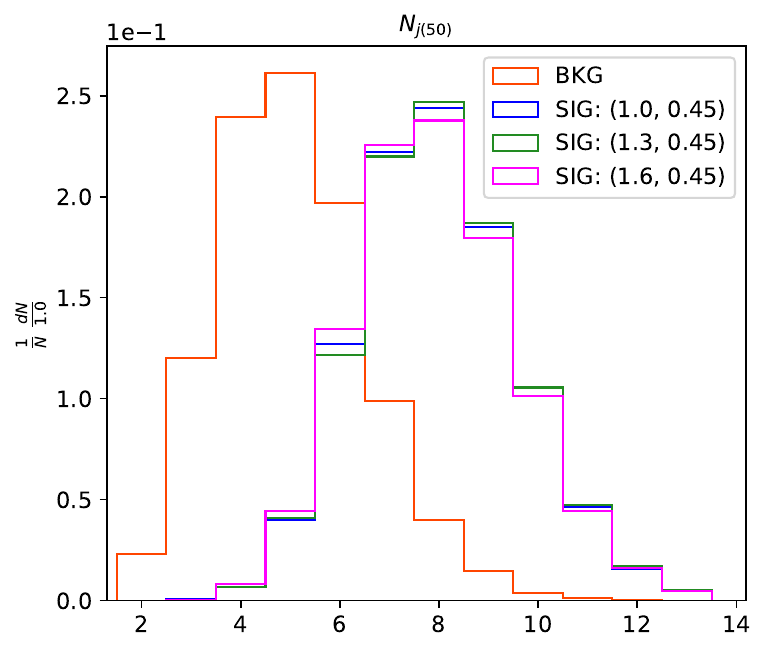}}\quad
\subfloat[]{\includegraphics[width=0.32\textwidth]{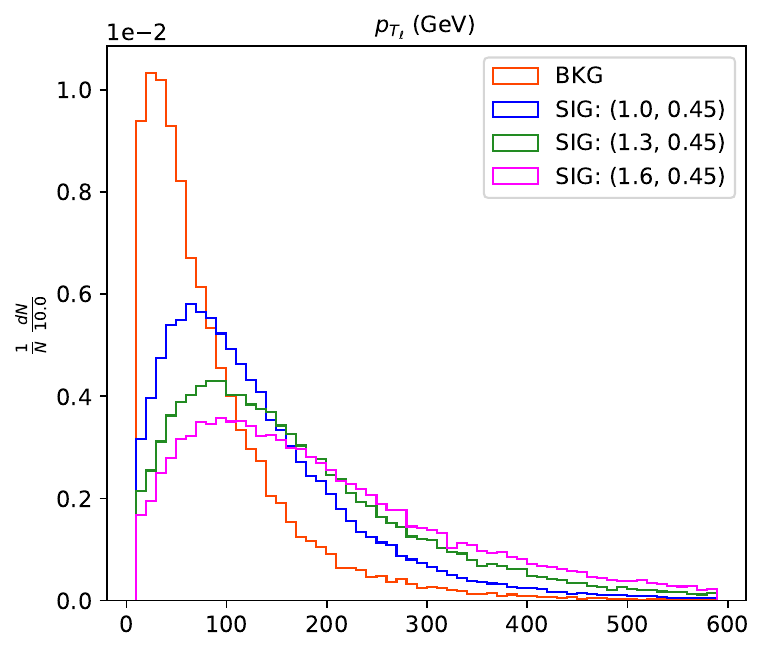}}\\
\subfloat[]{\includegraphics[width=0.32\textwidth]{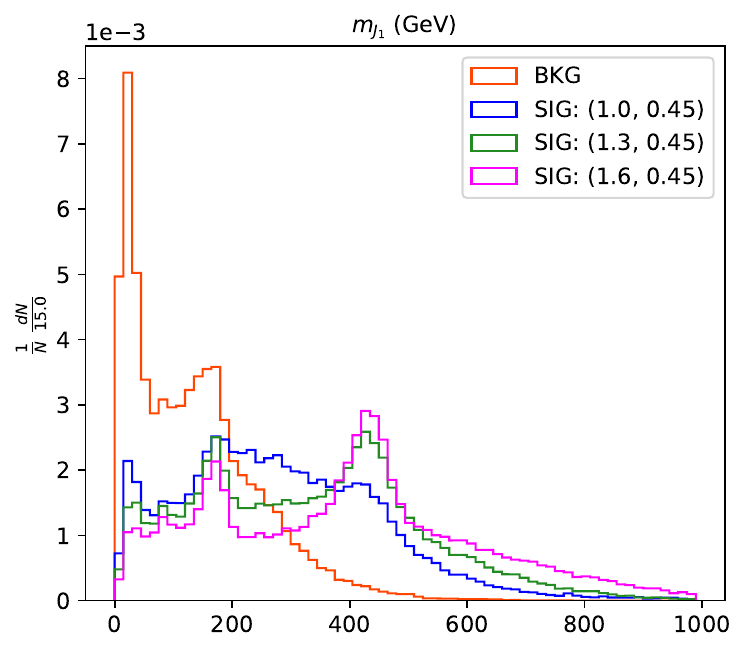}}\quad
\subfloat[]{\includegraphics[width=0.32\textwidth]{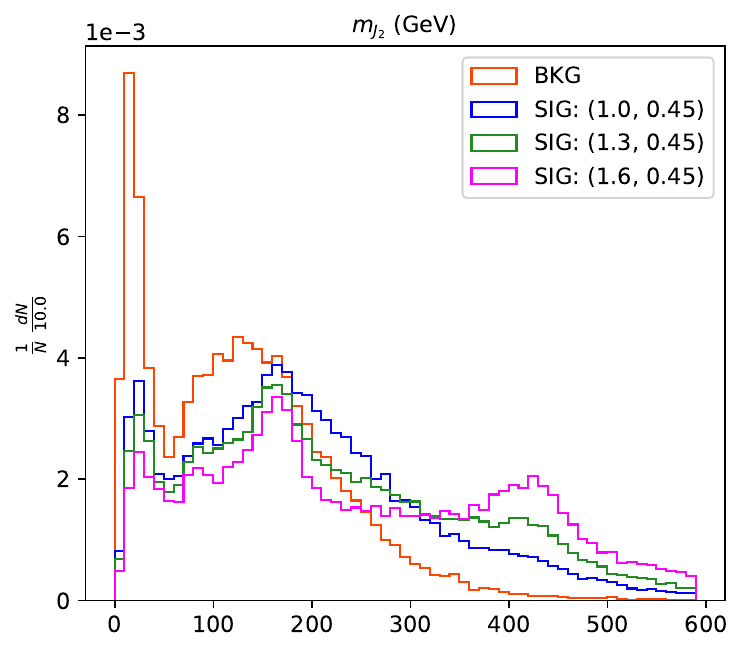}}\quad
\subfloat[]{\includegraphics[width=0.32\textwidth]{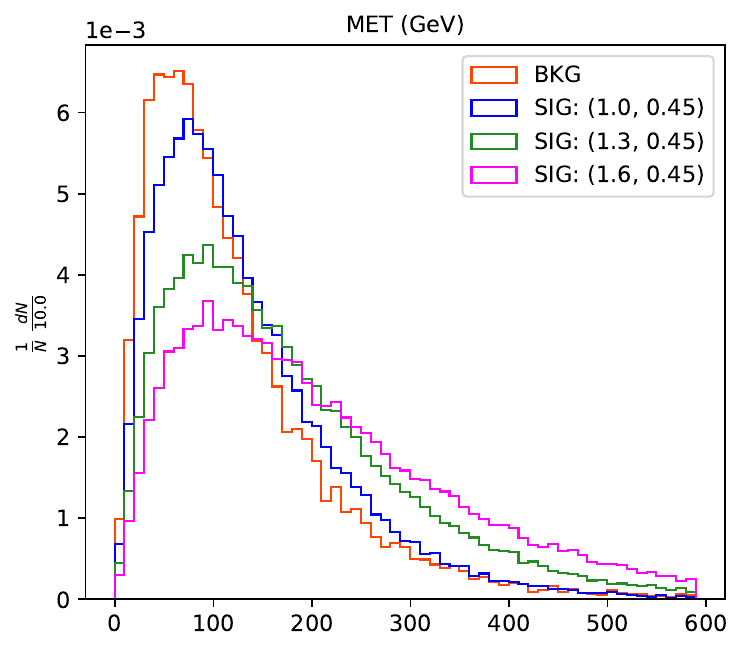}}\\
\subfloat[]{\includegraphics[width=0.32\textwidth]{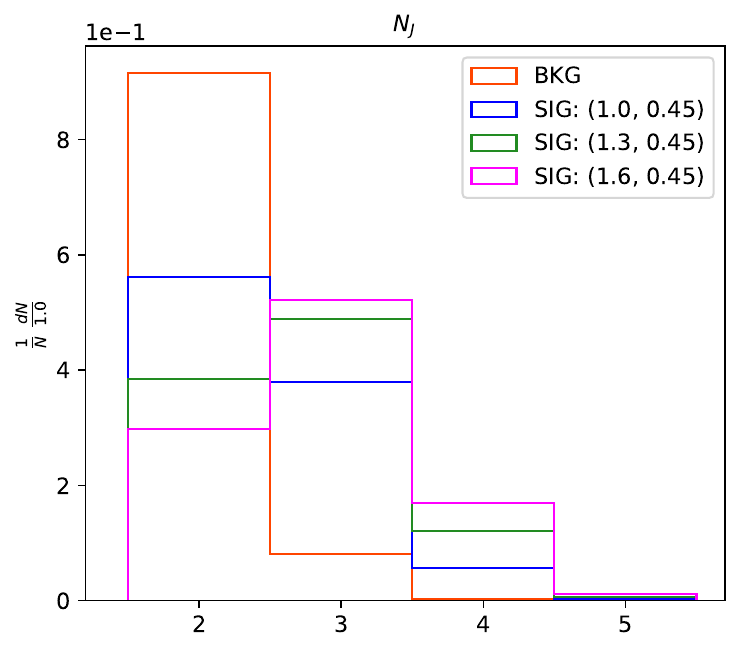}}\quad
\subfloat[]{\includegraphics[width=0.32\textwidth]{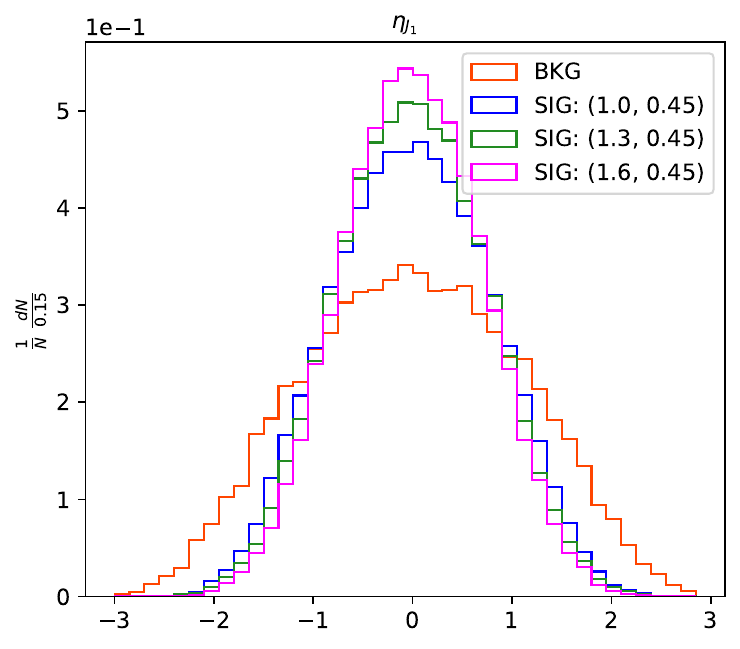}}\quad
\subfloat[]{\includegraphics[width=0.32\textwidth]{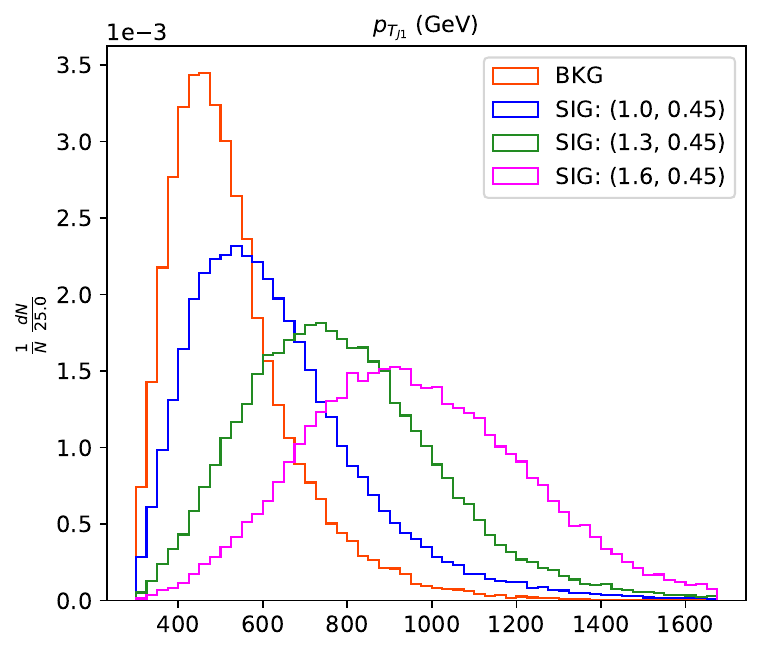}}
\caption{Normalised input feature distributions for signal and background at a fixed $M_{\Phi}=450$ GeV and for three different $M_{t_2}$ points: $M_{t_2}=1.0$, $1.3$ and $1.6$ TeV. The background distribution includes all the backgrounds processes mentioned in the text.}\label{fig:inputvars_mphi450}
\end{figure*}

We see some positive correlation between $\{H_{T},p_{T_{J1}}\}$ for both the signal and background. However, we retain both since $H_{T}$ is our highest performing variable and $p_{T_{J1}}$ is a good low-level fatjet feature. We also see some positive correlation between $\{p_{T_{J1}},m_{J_1}\}$ only for the signal. Since this correlation is not present for the background, we retain $m_{J_1}$ in our analysis.  

We show the input feature distributions for $M_{\Phi}=150$ and $450$ GeV in Figs.~\ref{fig:inputvars_mphi150} and~\ref{fig:inputvars_mphi450}, respectively. 
In these plots, the background distributions are obtained by a weighted sum of the separate processes:
\begin{equation}
    H = \frac{\sum_{i} h_{i} w_{i}}{\sum_{i} w_{i} }
\end{equation}
where $w_{i}=N_{\text{exp}}\slash N_{\text{gen}}$ ($N_{\text{exp}}$ and $N_{\text{gen}}$ are the expected number of events at $3$ ab$^{-1}$ and the number of generated events passing the selection criteria, respectively). The variable $H_{T}$ is driven by the mass of $t_2$---a heavier $t_2$ imparts more momentum to the AK4 jets and consequently pushes $H_{T}$ to higher values. A heavier $\Phi$ takes a larger fraction of the momentum of the parent $t_2$, leaving a smaller budget for the top quark, and as a consequence, the lepton and neutrino from the top decay are less boosted. This can be seen in the distributions for $p_{T_\ell}$ and MET as they are slightly pushed towards the lower values for $M_{\Phi}=450$ GeV than the $M_{\Phi}=150$ GeV case. The $\eta_{J_1}$ and $p_{T_{J1}}$ features are more or less similar for the two $\Phi$ masses as they merely describe the kinematics of the leading fatjet in the event, be it a top jet or a $gg$ fatjet from a $\Phi$.

While most of the feature distributions look similar for $M_{\Phi}=150$ GeV and $M_{\Phi}=450$ GeV, some of them do show differences. Since the $\Phi$'s originate in $t_2$ decays, for a fixed $M_{t_2}$, a heavier $\Phi$ is less boosted than a lighter one. As a result, the two subjets from $\Phi$ are more separated for heavy scalars, leading to a shift in the number of AK4 jets and a lower clustering efficiency for $R=1.2$ fatjets. This in turn affects $N_{J}$, $m_{J_1}$ and $m_{J_2}$ (see Fig.~\ref{fig:inputvars_mphi450}).

\subsection{BDT results}\label{sec:results}
\noindent
We optimise the statistical significance of the signal in  the BDT analysis. In absence of any systematic uncertainty, the signal significance over the background can be estimated by the approximate Poisson significance for the Asimov dataset:
\begin{equation}
    \mathcal{Z}=\sqrt{2\left(N_{S}+N_{B}\right)\ln\left(1+\frac{N_{S}}{N_{B}}\right)-2N_{S}}\label{eq:asimov}
\end{equation}
where $N_{S}$ and $N_{B}$ are the number of signal and background events after the optimal cut is applied on the BDT response. For $N_{S}\ll N_{B}$, Eq.~\eqref{eq:asimov} reduces to:
\begin{equation}
    \mathcal{Z}=\frac{N_{S}}{\sqrt{N_{S}+N_{B}}}.
\end{equation}
The above expression is our definition of statistical significance.

We optimise the BDT hyperparameters with the Adaptive Boosting algorithm~\cite{adaboost_ref} at the benchmark mass point: $\left(M_{t_2},M_{\Phi}\right)=\left(1.30 \text{ TeV}, 0.35 \text{ TeV}\right)$ which is roughly in the middle of the mass region considered. We split the dataset into statistically independent subsets for training and testing in a $50$-$50$ ratio. The criteria for optimising the BDT hyperparameters are (i) Kolmogorov-Smirnov test values between $0.1$ and $0.9$ for both signal and background, and (ii) a smooth significance curve. A summary of the optimised BDT parameters is given in Table~\ref{table:bdt_params}. In Table~\ref{table:feature_sepn}, we show the method-specific and method-unspecific rankings for the optimised BDT hyperparameters at the benchmark point. Defined in Ref.~\cite{Hocker:2007ht}, the input features are ranked by how often they are used to split nodes in decision trees, weighted by the number of events in the node and the separation gain squared achieved at the node (in our case it is the Gini-Index squared).

\begin{table}[!t]
\begin{centering}
\begin{tabular*}{\columnwidth}{c @{\extracolsep{\fill}} c}
\hline 
BDT parameter & Optimised choice\tabularnewline
\hline 
\hline 
NTrees & $300$\tabularnewline
MinNodeSize & $5.0$\%\tabularnewline
MaxDepth & $3$\tabularnewline
BoostType & AdaBoost\tabularnewline
AdaBoostBeta & $0.07$\tabularnewline
UseBaggedBoost & True\tabularnewline
BaggedSampleFraction & $0.5$\tabularnewline
SeparationType & GiniIndex\tabularnewline
nCuts & 50\tabularnewline
\hline 
\end{tabular*}
\par\end{centering}
\caption{Summary of optimised BDT hyperparameters.}\label{table:bdt_params}
\end{table}
\begin{table}
\begin{centering}
\begin{tabular*}{\columnwidth}{c @{\extracolsep{\fill}} ccc}
\hline 
\multirow{2}{*}{Feature} & Method Unspecific & Feature & Method Specific\tabularnewline
 & Separation &  & Ranking\tabularnewline
\hline 
\hline 
$H_{T}$ & $0.6689$ & $H_{T}$ & $0.3606$\tabularnewline
$p_{T_{J1}}$ & $0.4326$ & $N_{j\left(50\right)}$ & $0.1663$\tabularnewline
$N_{j\left(50\right)}$ & $0.4241$ & $p_{T_\ell}$ & $0.1205$\tabularnewline
$m_{J_{1}}$ & $0.3220$ & $m_{J_{1}}$ & $0.0923$\tabularnewline
$N_{J}$ & $0.2913$ & $m_{J_{2}}$ & $0.0706$\tabularnewline
$p_{T_\ell}$ & $0.2104$ & $\text{MET}$ & $0.0598$\tabularnewline
$m_{J_{2}}$ & $0.1608$ & $N_{J}$ & $0.0585$\tabularnewline
$\text{MET}$ & $0.1068$ & $\eta_{J1}$ & $0.0406$\tabularnewline
$\eta_{J1}$ & $0.0722$ & $p_{T_{J1}}$ & $0.0305$\tabularnewline
\hline 
\end{tabular*}
\par\end{centering}
\caption{Summary of method unspecific and method-specific ranking for the input features at the benchmark mass point.}\label{table:feature_sepn}
\end{table}

\begin{figure*}[t]
\begin{centering}

\subfloat[]{\includegraphics[width=0.48\textwidth]{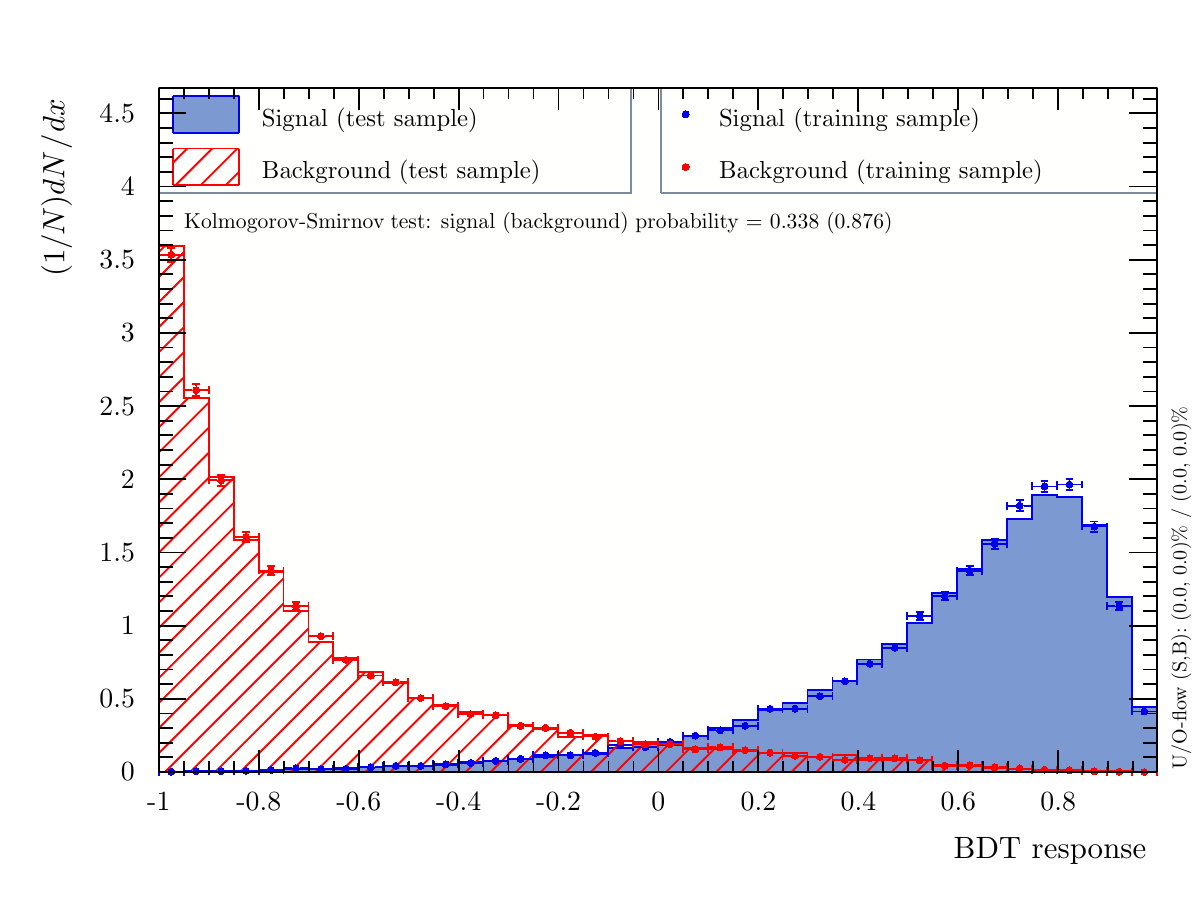}}\hspace{0.2cm}
\subfloat[]{\includegraphics[width=0.48\textwidth]{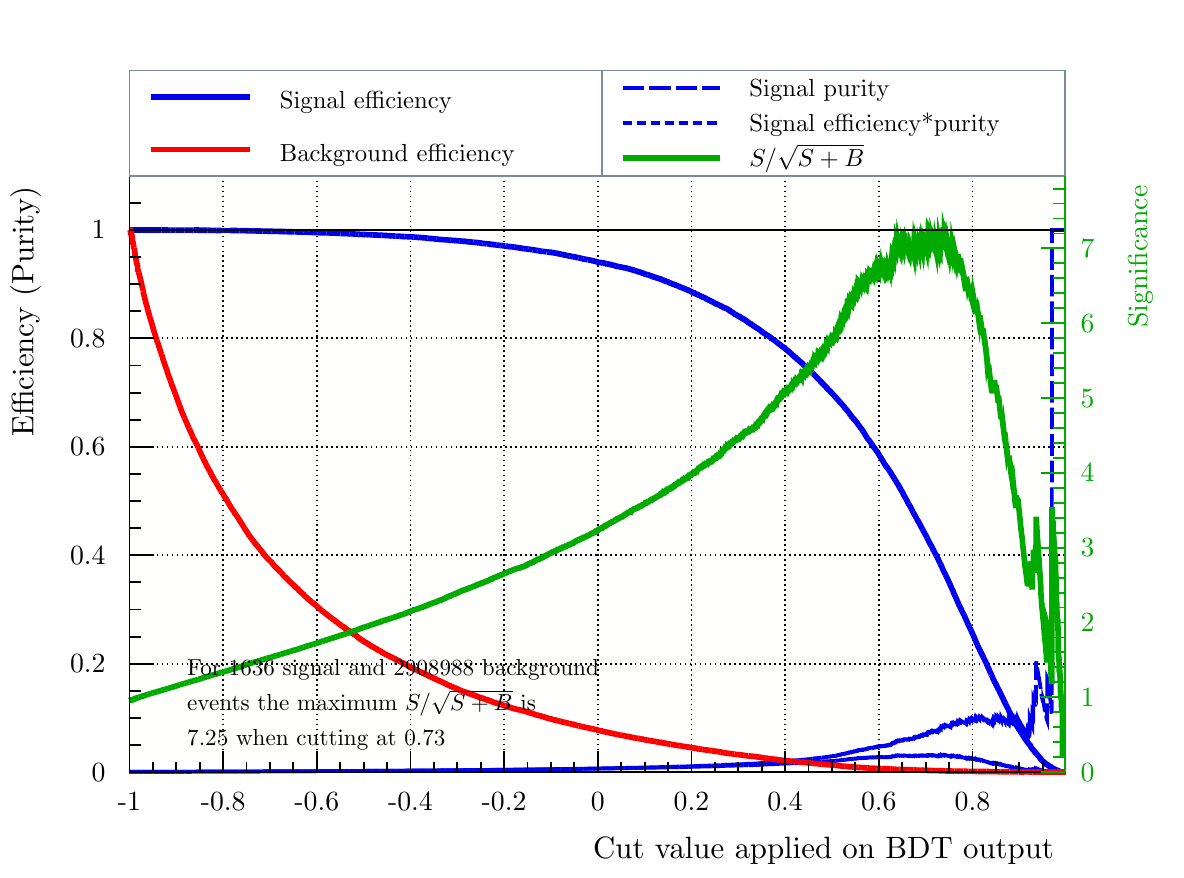}}
\par\end{centering}
\caption{The BDT response curve and overtraining check is shown in (a) and the cut efficiencies and $\mathcal{Z}$ value at the optimal BDT cut are shown in (b).}\label{fig:bdt_results}
\end{figure*}
\begin{figure}[t]
\includegraphics[width=0.95\columnwidth]{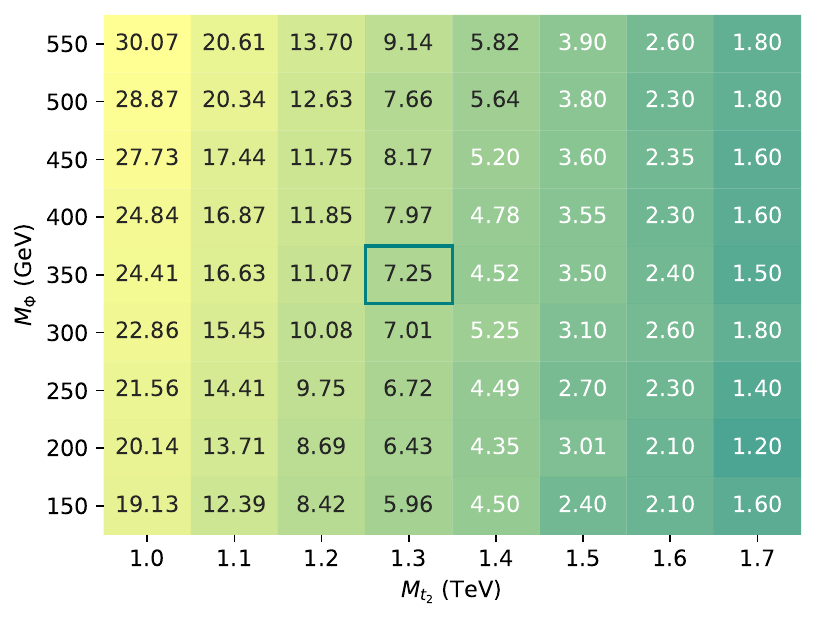}
\caption{Statistical significance $\mathcal{Z}$ over the mass range considered. Benchmark mass is highlighted.}\label{fig:final_signif}
\end{figure}
\begin{figure*}
\captionsetup[subfigure]{labelformat=empty}
\begin{centering}
\subfloat[\quad\quad(a)]{\includegraphics[width=0.44\textwidth]{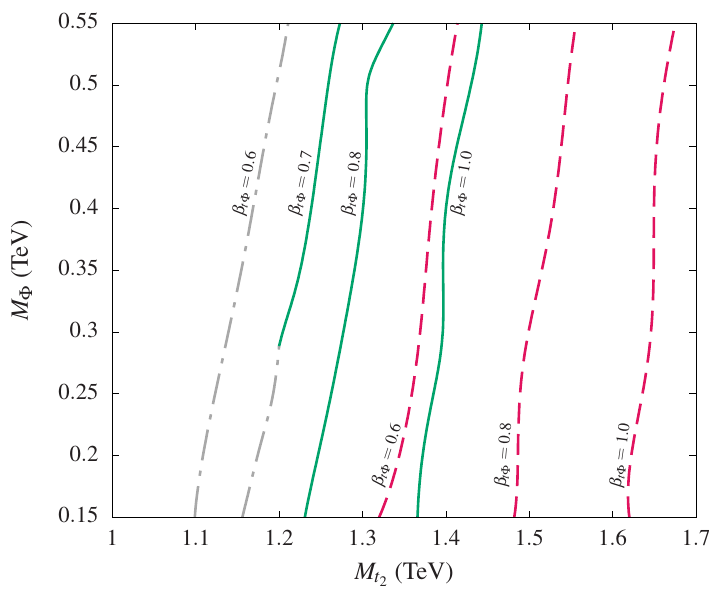}}
\hspace{1cm}\subfloat[\quad\quad(b)]{
\includegraphics[width=0.44\textwidth]{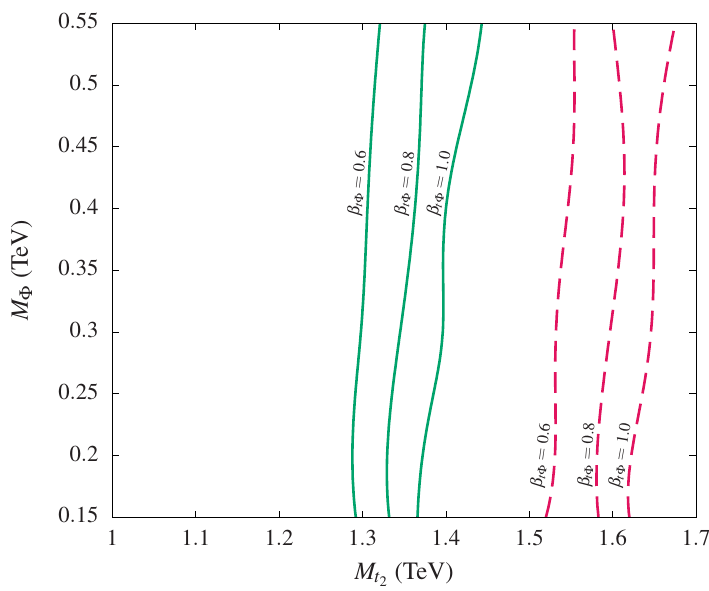}
}
\par\end{centering}
\caption{Discovery reach for $\text{BR}\left(T\rightarrow t \Phi\right)\leq1$ for (a) the exclusive $(t\Phi) (\bar{t}\Phi)$ mode and (b) the inclusive $t\Phi + X$ mode. The $5\sigma$ discovery contours are drawn in green (solid lines) and the $2\sigma$ exclusion contours are in red (dashed lines), marked with the corresponding branching ratios. The parts marked in grey (dot-dashed lines) are excluded by the LHC data as shown in Fig.~\ref{fig:t2_LHClimits}.}\label{fig:reach_diff_modes}
\end{figure*}

We use the hyperparameters optimised at the benchmark mass point to obtain the statistical significance $\mathcal{Z}$ with the optimal BDT cut at all mass points. For $M_{t_2}>1.5$ TeV, where the signal cross sections are low, we use the $\mathcal{Z}$ value obtained for a slightly relaxed cut on the BDT response. We show the BDT response and cut efficiency at the benchmark mass point in Fig.~\ref{fig:bdt_results}. We list $\mathcal{Z}$ at each point in the mass parameter scan in Fig.~\ref{fig:final_signif}. We see that, unlike the selection efficiency, the $5\sigma$ and $2\sigma$ contours follow approximately constant $M_{t_2}$ lines near $M_{t_2}=1.4$ TeV and $M_{t_2}=1.7$ TeV, respectively. For a fixed $M_{t_2},$ the decay products of a heavy $\Phi$ are more separated than that of a lighter $\Phi$, so the clustering algorithm is less likely to pick up the entire $gg$ fatjet. Hence, the fatjet features become less separated between the signal and background distributions, leading to a drop in significance. As a result, the gain in the selection efficiency for a heavier $\Phi$ (see Figure \ref{fig:signal_cuteff}) is lost to a slight degree in the multivariate analysis (MVA).

\subsection{Inclusive signal}\label{subsec:inclsig}
\noindent
With a MVA optimised for the $\Phi t \Phi \bar{t}$ channel with $\bt_{t\Phi}=1$, we explore the discovery reach for more realistic scenarios where $\text{BR}\left(T\rightarrow t \Phi\right)<1$. Since the signal scales as $\beta_{t\Phi}^{2}$ in the exclusive mode, its significance drops very fast with a decreasing $\beta_{t\Phi}$ (see Fig.~\ref{fig:reach_diff_modes}).
However, for $\beta_{t\Phi}<1$, when other $t_2$ decay modes open up, they can be added to the signal definition. If we consider the mixed mode, i.e., consider 
\begin{align}
   pp\to t_{2}\bar{t}_{2}\to (t\Phi)(\bar{t}\Phi/\bar{b}W/\bar{t} Z/\bar{t} h)+\text{c.c.}\label{eqn:mixed_modes}
\end{align}
as our signal, a significant fraction of events from all other decay modes would also pass the selection criteria (Sec.~\ref{sec:sigtop}) because of their inclusive nature in addition to the $t\Phi$ events. We analyse this case for some representative branching ratios. The mixed mode events generated for $\bt_{t\Phi}<1$ are passed through the selection cuts and are fed to the optimised MVA. We plot the $3$ ab$^{-1}$ $5\sg$ (discovery) and $2\sg$ (exclusion) contours with the three values of $\bt_{t\Phi}$ for both the exclusive and inclusive signals in Fig.~\ref{fig:reach_diff_modes}. We see that 
the discovery and exclusion reaches in the inclusive mode are significantly higher than those in the exclusive mode. From the figure, we see that a discovery in the exclusive mode is not possible if $\bt_{t\Phi}\leq 0.6$ as the region is excluded by the LHC limits from Fig.~\ref{fig:t2_LHClimits}, whereas for $\bt_{t\Phi} \sim 0.7$, it is possible if $M_\Phi \gtrsim 300$ GeV. In the inclusive mode, $\beta_{t\Phi} \sim 0.6$ is the lower limit for a discovery.

\subsection{Single production}\label{subsec:single}
\noindent 
We also analyse the single production of $t_2$. In the absence of the top quark in the initial state, producing it singly at the LHC is tough. The dominant process is $pp\rightarrow t_2 {b} j$ through a $t$-channel $W$ boson exchange. We generate the process and decay the VLQ as $t_2\rightarrow t \Phi \rightarrow \left(l\nu b\right)\left(gg\right)$. Since a single production is less phase space suppressed than the pair production, its cross section falls slower than that of the pair production with increasing $M_{t_2}$. Therefore, one might na\"ively expect the single production to have a better discovery potential in the high VLQ mass region. However, two factors go against it. There is a competition between the mixing angle $\theta_{L}$ and $\bt_{t\Phi}$---it is difficult to obtain a large single-production cross section (the $t_2bW$ coupling goes as $\sin\theta_L$ times the $tbW$ coupling~\cite{Bhardwaj:2022nko}) and a significant BR in the $t_2\to t \Phi$ mode simultaneously. Second, the single production signal suffers from a low selection efficiency due to a relatively lower $H_{T}$ (as a consequence of only one $T$ being present in the process) and the absence of boosted fatjets. The $H_{T}$ distribution for single production looks very similar to the background, so its discriminating power is also lower than the $\Phi t\Phi t$ channel. We have tested these by generating the single production signal with $\sin\theta_L\sim0.1$. Overall, we find that with the $t_2$ decaying to the singlet $\Phi$, the single production of $t_2$ is not a promising channel.

\section{Summary and Conclusions}\label{sec:conclusion}
\noindent 
In this paper, we have analysed the pair production of a heavy weak-singlet vectorlike top partner and its subsequent decay into a top quark and a weak-singlet (pseudo)scalar $\Phi$, which gives a di-jet signature. In an earlier paper~\cite{Bhardwaj:2022nko}, we had shown that the available parameter space in a singlet $T$ model, where the singlet $\Phi$ dominantly decays into two gluons, is pretty large and that the collider searches specific to this signature can be used to probe the $(M_{t_2}, M_{\Phi})$ parameter space effectively. However, the signal for this signature is particularly challenging to isolate from a formidable SM background. We have made use of multivariate analysis techniques, specifically BDTs. We have also utilised jet substructure techniques to tag $\Phi$ as a two-prong structure; these features in the signal provided an extra handle to control the SM backgrounds. We optimised the analysis for an ideal scenario where BR$(t_2 \to t \Phi) = 1.0$ and BR$(\Phi \to g g) = 1.0$ and then presented some realistic cases for the $t_2 \to t\Phi$ branching which can be probed at the LHC. We have shown the discovery and exclusion prospects for various combinations of $t_2$ and $\Phi$ masses in both exclusive ($t_2 \bar{t}_2 \to t\Phi \bar{t} \Phi$) as well as inclusive ($t_2 \bar{t}_2 \to t\Phi + X$) modes, where $X$ is any of the standard decay modes of $t_2$. 

There are other similar extensions to the SM containing heavy top partners in different weak representations (e.g., doublet) with  a new decay mode $(t_2 \to t\Phi)$. Since we treat the branching ratios as independent quantities, our search strategy allows for straightforward projection of limits in these models as well. Our study also offers predictive insights into exclusive and inclusive VLQ searches at the LHC.

\section*{Model Files}
\noindent 
The {\sc Universal FeynRules Output}~\cite{Degrande:2011ua} files used in this paper are available at \url{https://github.com/rsrchtsm/vectorlikequarks/} under the name {\tt SingTplusPhi}. 

\acknowledgments
\noindent
We acknowledge the high-performance computing time at the Padmanabha cluster, IISER Thiruvananthapuram, India.  A. B. is supported by the STFC under Grant No. ST/T$000945$/$1$. K. B. and C. N. acknowledge DST-Inspire for their fellowships.

\begin{figure*}
\centering
\captionsetup[subfigure]{labelformat=empty}
\subfloat[\quad\quad(a)]{\includegraphics[width=0.8\columnwidth]{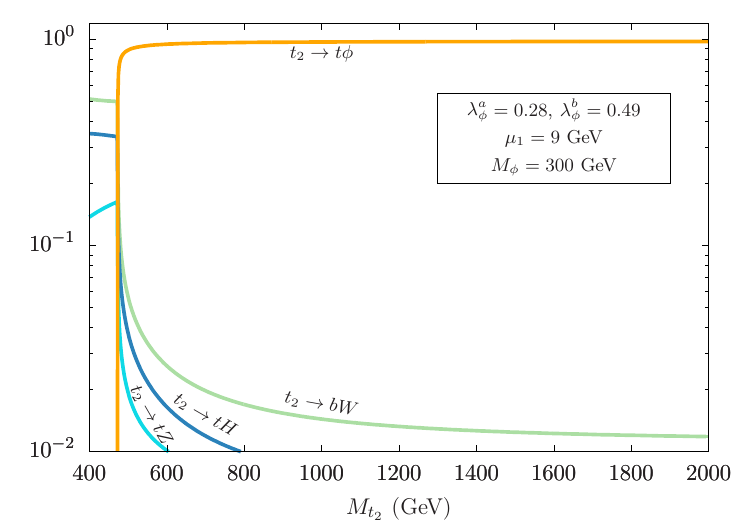}\label{fig:T2_bMark1}}\hspace{2cm}
\subfloat[\quad\quad(b)]{\includegraphics[width=0.8\columnwidth]{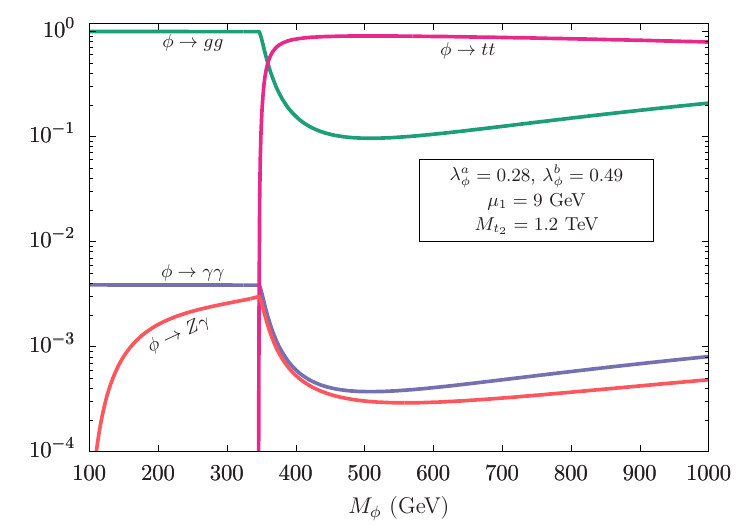}\label{fig:Phi_bMark1}}\\
\subfloat[\quad\quad(c)]{\includegraphics[width=0.8\columnwidth]{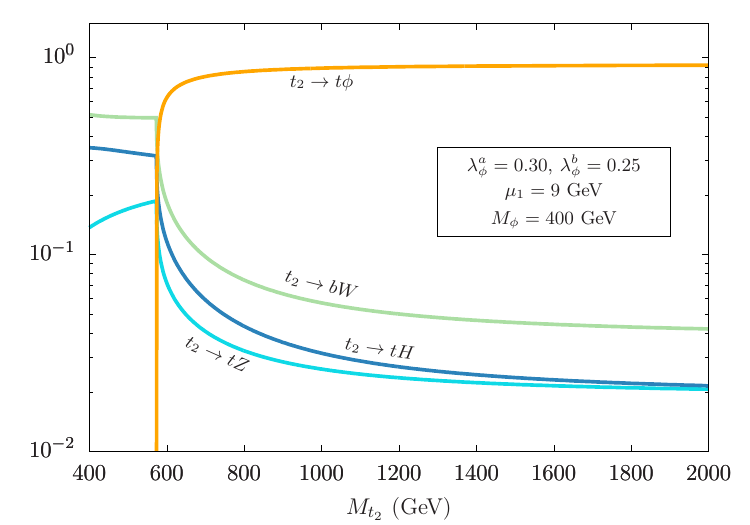}\label{fig:T2_bMark2}}\hspace{2cm}
\subfloat[\quad\quad(d)]{\includegraphics[width=0.8\columnwidth]{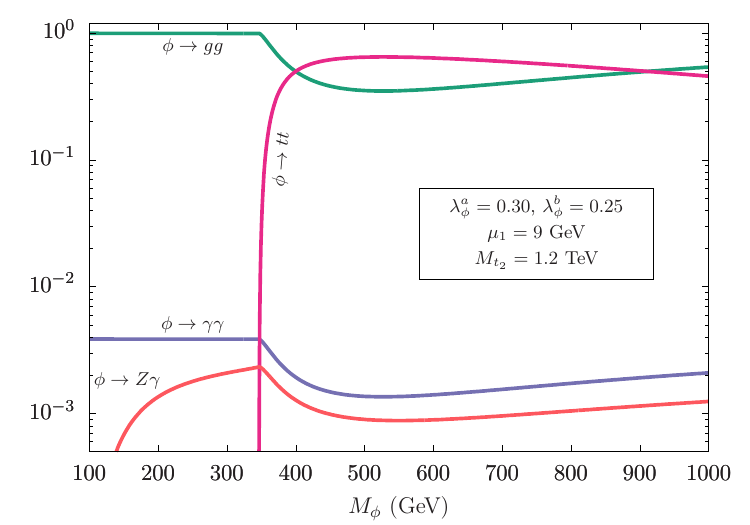}\label{fig:Phi_bMark2}}\\
\subfloat[\quad\quad(e)]{\includegraphics[width=0.8\columnwidth]{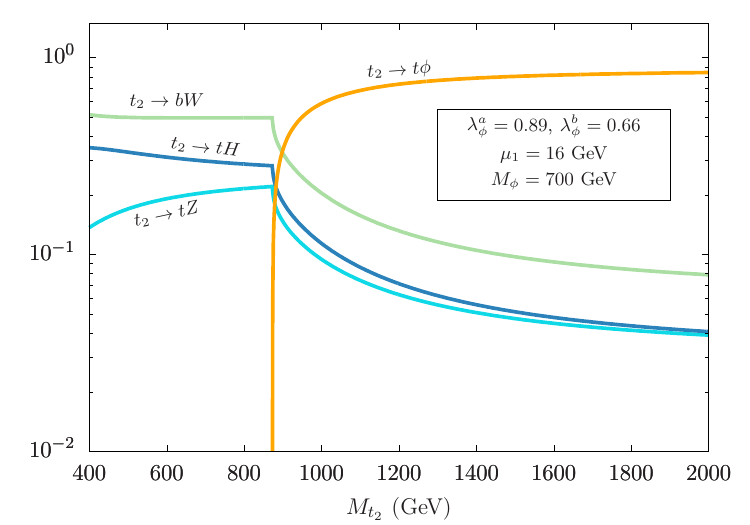}\label{fig:T2_bMark3}}\hspace{2cm}
\subfloat[\quad\quad(f)]{\includegraphics[width=0.8\columnwidth]{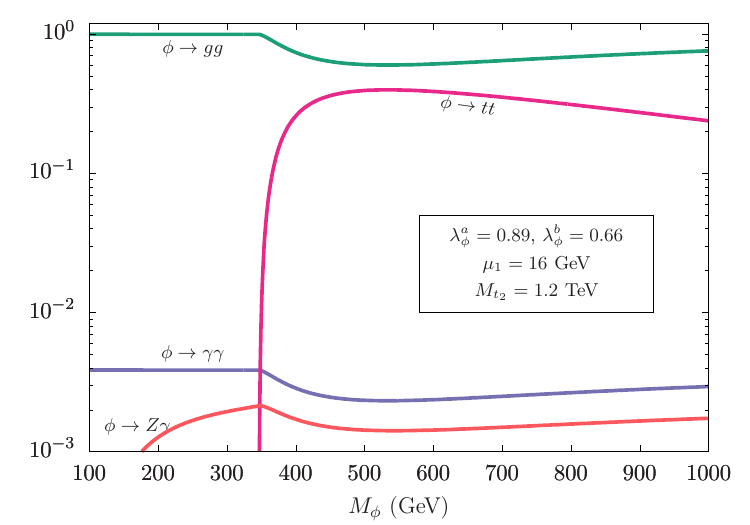}\label{fig:Phi_bMark3}}\\
\subfloat[\quad\quad(g)]{\includegraphics[width=0.8\columnwidth]{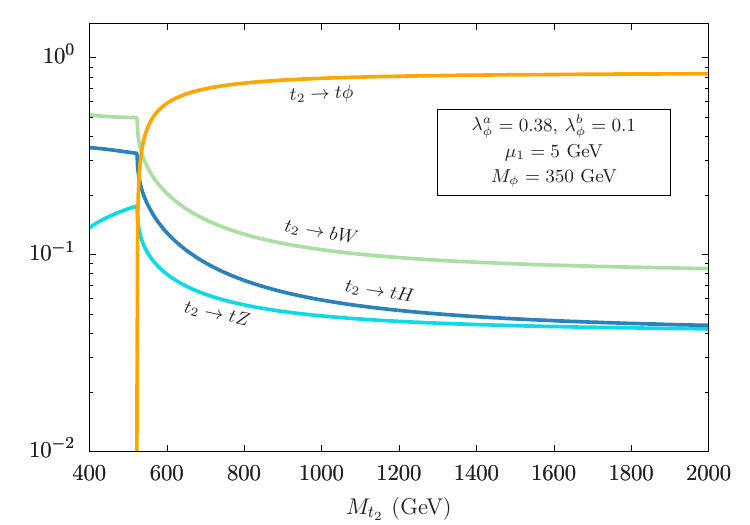}\label{fig:T2_bMark4}}\hspace{2cm}
\subfloat[\quad\quad(h)]{\includegraphics[width=0.8\columnwidth]{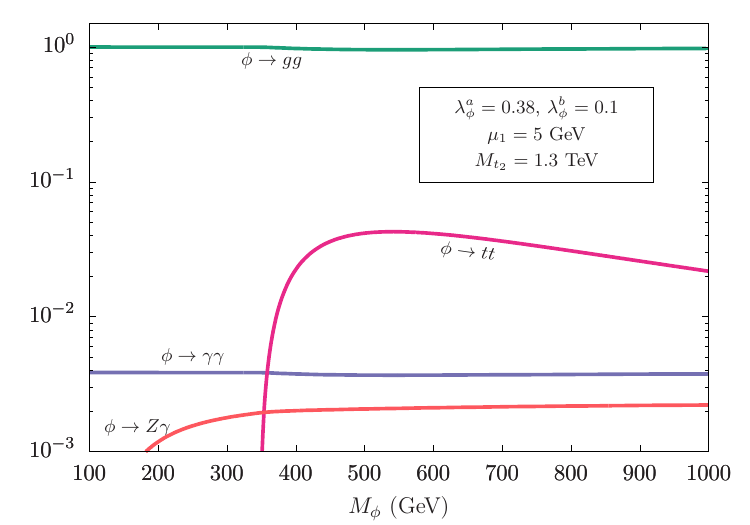}\label{fig:Phi_bMark4}}
\caption{Illustrative BR plots for $t_2$ and $\Phi$ decays. Panels (a)-(f) relate to the benchmark parameters identified in Ref.~\cite{Bhardwaj:2022nko} for the singlet $T$ model. Panels (g) and (h) correspond to the parameters we have used to optimise the analysis in this paper.}
\label{fig:BR}
\end{figure*}

\appendix
\section{Lagrangian and Benchmarks}
\label{app:mod}

\noindent We reproduce the terms relevant for the top sector mass matrix from Ref.~\cite{Bhardwaj:2022nko} as,
\begin{align*}
\mc{L} \supset -\Big\{&\lm_{t}\lt(\bar{Q}_LH_T\rt)t_R
+\om_{T}\lt(\bar{Q}_LH_T\rt)T_R\nn\\
& + M_{T}\bar{T}_LT_R+ {\tr h.c.}\Big\},
\end{align*}
where $Q_L$ is the third-generation left-handed quark doublet and $H_T = i\sigma_2H^*$, with $H$ being the Higgs doublet. The top Yukawa coupling is denoted by 
$\lm_t$, $\omega_T$ is the off-diagonal coupling and $M_T$ is the vectorlike mass term.  After EWSB, we get the mass matrix $\mc M$ as,  
\begin{align*}
\mc{L}_{mass} =& \bpm \bar{t}_L & \bar{T}_L \epm 
\bpm 
\begin{array}{cc}
\lm_t \frac v{\sqrt2} & \l\om_{T}\,\frac v{\sqrt2} \\ 0 & M_T
\end{array} 
\epm 
\bpm t_R \\ T_R \epm + {\tr h.c.},
\end{align*}
where $v$ is the Higgs vacuum expectation value. The  matrix $\mc M$ can be diagonalised by bi-orthogonal rotations and the left and right mixing angles are given by,
\begin{align*}
\tan{(2\theta_L)} =  \frac{2M_T\,\m_{1}}{m_t^2+\m_{1}^2-M_T^2},\ 
\tan{(2\theta_R)} =  \frac{2m_t\,\m_{1}}{m_t^2-M_T^2-\m_{1}^2}.
\end{align*}
where $m_t = \lm_t \frac v{\sqrt2}$ and $\mu_1 = \omega_T \frac v{\sqrt2}$. The mass eigenvalues $m_{t_1}$ and $M_{t_2}$ are 
\begin{align*}
m_{t_1}^2,M_{t_2}^2 &=  \frac12\Bigg[\lt(m_t^2+\m_1^2+M_T^2\rt) \nn\\ 
&\quad\quad\quad\mp\sqrt{\lt(m_t^2+\m_1^2+M_T^2\rt)-4m_t^2M_T^2}\Bigg].
\end{align*}
Expressions for the partial decay widths of $t_2$ and $\Phi$ are found in Ref.~\cite{Bhardwaj:2022nko}. With those expressions, one can obtain the BRs of these particles. 
The BR plots for some benchmark points are shown in Fig.~\ref{fig:BR}---Figs.~\ref{fig:T2_bMark1}--~\ref{fig:Phi_bMark3} relate to the benchmarks presented in Ref.~\cite{Bhardwaj:2022nko} for the singlet $T$ model with a scalar $\phi$. Figs.~\ref{fig:T2_bMark4} and \ref{fig:Phi_bMark4} show the behaviour of the benchmark point $\{M_{t_2}, M_\phi\} = \{1.3, 0.35\}$ TeV for which the analysis is optimised. All the plots show scenarios in which the $t_2 \to t\Phi$ mode dominates. From the exclusion plot in Fig.~\ref{fig:t2_LHClimits}, we see that only beyond $M_{t_2}=1.3$ TeV, the standard decays of $t_2$ can dominate, i.e., BR$(t_2\to t\phi)\lesssim 0.5$. For example, we get BR$(t_2\to t\phi)\approx 0.5$ for $M_{t_2}=1.3$ TeV, $\lambda^a_\phi \approx 0.26$, $\lambda^b_\phi \approx 0.1$, $\mu_1 \approx 10 $~GeV and $M_\phi \approx 350$~GeV. Hence,  if the couplings are smaller for the same mass parameters (or if $\m_1$ is bigger with the other parameters held fixed), the standard decays of the singlet top partner will dominate over the new mode.



\bibliography{bibliography.bib}
\bibliographystyle{JHEPCust}

\end{document}